\documentclass[aps,prl,twocolumn,superscriptaddress,10pt,noshowpacs,floatfix]{revtex4-1}
\usepackage[pdftex,plainpages=false,colorlinks=true,citecolor=blue,linkcolor=blue,urlcolor=blue,filecolor=green,bookmarksopen=true]{hyperref}
\usepackage[utf8]{inputenc}
\usepackage{amsmath}
\usepackage[english]{babel}
\usepackage[normalem]{ulem}
\usepackage[caption = false]{subfig}
\usepackage{graphicx,epstopdf}
\usepackage{blindtext}
\usepackage{lipsum}
\usepackage{amsfonts}
\usepackage{bbm}
\usepackage{amssymb}
\usepackage{enumerate}
\usepackage{color}
\usepackage[dvipsnames]{xcolor}
\usepackage{latexsym}
\usepackage{times,txfonts}

\newcommand{\Hcal}{\mathcal{H}}
\newcommand{\Dcal}{\mathcal{D}}
\newcommand{\Pcal}{\mathcal{P}}

\newcommand{\Lcal}{\mathcal{L}}
\newcommand{\Ecal}{\mathcal{E}}
\newcommand{\Rcal}{\mathcal{R}}

\newcommand{\Zcal}{\mathcal{Z}}
\newcommand{\Fcal}{\mathcal{F}}

\newcommand{\1}{\mathbbm{1}}
\newcommand{\Lmath}{\mathbbm{L}}
\newcommand{\Rmath}{\mathbbm{R}}
\newcommand{\Hmath}{\mathbbm{H}}

\newcommand{\dket}[1]{| #1 \rangle\rangle}
\newcommand{\dbra}[1]{\langle\langle #1 |}
\newcommand{\ket}[1]{| #1 \rangle}
\newcommand{\bra}[1]{\langle #1 |}

\newcommand{\dinterpro}[2]{\langle \langle #1 | #2 \rangle \rangle}

\newcommand{\tr}[1]{ \text{Tr}\left\{ #1 \right\}}
\newcommand{\trs}[1]{ \text{Tr} \{ #1 \}}

\DeclareFontFamily{U}{mathc}{}
\DeclareFontShape{U}{mathc}{m}{it}%
{<->s*[1.03] mathc10}{}
\DeclareMathAlphabet{\mathscr}{U}{mathc}{m}{it}

\begin{document}

\title{Quantum thermodynamics in adiabatic open systems and its trapped-ion experimental realization}

\author{Chang-Kang Hu}
\affiliation{CAS Key Laboratory of Quantum Information, University of Science and Technology of China, Hefei 230026, People’s Republic of China}
\affiliation{Shenzhen Institute for Quantum Science and Engineering, Southern University of Science and Technology, Shenzhen 518055, China}
\affiliation{CAS Center For Excellence in Quantum Information and Quantum Physics, University of Science and Technology of China, Hefei 230026, People’s Republic of China}

\author{Alan C. Santos}
\email{ac\_santos@id.uff.br}
\affiliation{Instituto de F\'{i}sica, Universidade Federal Fluminense, Av. Gal. Milton Tavares de Souza s/n, Gragoat\'{a}, 24210-346 Niter\'{o}i, Rio de Janeiro, Brazil}

\author{Jin-Ming Cui}
\email{jmcui@ustc.edu.cn}
\affiliation{CAS Key Laboratory of Quantum Information, University of Science and Technology of China, Hefei 230026, People’s Republic of China}
\affiliation{CAS Center For Excellence in Quantum Information and Quantum Physics, University of Science and Technology of China, Hefei 230026, People’s Republic of China}

\author{Yun-Feng Huang}
\email{hyf@ustc.edu.cn}
\affiliation{CAS Key Laboratory of Quantum Information, University of Science and Technology of China, Hefei 230026, People’s Republic of China}
\affiliation{CAS Center For Excellence in Quantum Information and Quantum Physics, University of Science and Technology of China, Hefei 230026, People’s Republic of China}

\author{\\Diogo O. Soares-Pinto}
\email{dosp@ifsc.usp.br}
\affiliation{Instituto de Física de São Carlos, Universidade de São Paulo, CP 369, 13560-970, São Carlos, SP, Brazil}

\author{Marcelo S. Sarandy}
\email{msarandy@id.uff.br}
\affiliation{Instituto de F\'{i}sica, Universidade Federal Fluminense, Av. Gal. Milton Tavares de Souza s/n, Gragoat\'{a}, 24210-346 Niter\'{o}i, Rio de Janeiro, Brazil}

\author{Chuan-Feng Li}
\email{cfli@ustc.edu.cn}
\affiliation{CAS Key Laboratory of Quantum Information, University of Science and Technology of China, Hefei 230026, People’s Republic of China}
\affiliation{CAS Center For Excellence in Quantum Information and Quantum Physics, University of Science and Technology of China, Hefei 230026, People’s Republic of China}

\author{Guang-Can Guo}
\affiliation{CAS Key Laboratory of Quantum Information, University of Science and Technology of China, Hefei 230026, People’s Republic of China}
\affiliation{CAS Center For Excellence in Quantum Information and Quantum Physics, University of Science and Technology of China, Hefei 230026, People’s Republic of China}

\begin{abstract}
Quantum thermodynamics aims at investigating both the emergence and the limits of the laws of thermodynamics from a quantum mechanical microscopic approach.
In this scenario, thermodynamic processes with no heat exchange, namely, adiabatic transformations, can be implemented through quantum evolutions
in closed systems, even though the notion of a closed system is always an idealization and approximation. Here, we begin by theoretically discussing
thermodynamic adiabatic processes in open quantum systems, which evolve non-unitarily under decoherence due to its interaction with its surrounding environment.
From a general approach for adiabatic non-unitary evolution, we establish heat and work in terms of the
underlying Liouville superoperator governing the quantum dynamics. As a consequence, we derive the conditions that an adiabatic open-system quantum
dynamics implies in the absence of heat exchange, providing a connection between quantum and thermal adiabaticity. Moreover, we determine
families of decohering systems exhibiting the same maximal heat exchange, which imply in classes of thermodynamic adiabaticity in open systems.
We then approach the problem experimentally using a hyperfine energy-level quantum bit of an Ytterbium $^{171}$Yb$^+$ trapped ion, which provides a work substance for thermodynamic processes, allowing for
the analysis of heat and internal energy throughout a controllable engineered dynamics.
\end{abstract}

\maketitle

\section*{Introduction} The notion of adiabaticity is a fundamental concept in a number of different areas in physics,
including quantum information processing~\cite{Farhi:01,Bacon:09,Bacon:17,Tameem:18} and quantum thermodynamics~\cite{Kieu:04,Maruyama:09,Abah:17}.
In the context of closed quantum systems, adiabaticity is understood as the phenomenon in which the Hilbert space of the system can be (quasi-)perfectly decomposed
into decoupled Schrodinger-eigenspaces, composed by the eigenvectors of the Hamiltonian with distinct non-crossing
instantaneous energies~\cite{Born:28,Kato:50,Messiah:Book}. Then, by initially preparing a quantum system in an energy eigenstate, the system undergoes a decoupled
evolution to the corresponding energy eigenstate at later times. However, the concept of a closed system is always an
idealization and approximation. Indeed, real quantum systems are always coupled to a surrounding environment.
In open quantum systems described by time-local master equations, the definition of adiabaticity can be naturally extended
to the decomposition of the Hilbert-Schmidt space into Lindblad-Jordan eigenspaces associated with distinct eigenvalues
of the generator of the dynamics~\cite{Sarandy:05-1,Sarandy:05-2,Sarandy:04,Venuti:16,Oreshkov:10,Patrik:05,Tameem:15}.

In thermodynamics, adiabaticity is associated to a process with no heat exchange
between the system and its reservoir. In general, it is not possible to associate an observable for the thermodynamic
definition of heat and of work~\cite{Talkner:07}. Then, the starting point widely used to define such physical quantities
in quantum systems is from the definition of internal energy given as $U(t) = \langle H(t) \rangle$~\cite{Alicki:79,Kieu:04}.
From this definition, we obtain the work ($d W$) and exchanged heat ($d Q$) between the reservoir and system as
\begin{eqnarray}
d W = \trs{ \rho (t) \dot{H}(t)}dt \mathrm{ \ \ \ \ and \ \ \ \ }d Q = \trs{ \dot{\rho} (t) H(t)}dt \label{delQ} \text{ , }
\label{WQori}
\end{eqnarray}
respectively. As originally introduced in Ref.~\cite{Alicki:79}, these quantities are defined in the weak coupling limit between 
system and reservoir (see also Refs.~\cite{Rivas:19,Guarnieri:20} for recent attempts to examine strongly coupled quantum systems and Refs.~\cite{AdolfoNewRef,SalimiNewRef}
for separation of internal energy variation
in terms of entropy changes). Notice also that $dW$ and 
$dQ$ are exact differential forms when at least one of them vanishes, thus the non-vanishing quantity can be
identified with the internal energy variation $\Delta U(t)$ during the entire process. For example, for a unitary transformation associated
with a closed quantum system, we necessarily have $d Q_{\mathrm{closed}}\!=\!0$, so that any variation $\Delta U(t)$ is due some work performed on/by the
system~\cite{Kieu:04,Anders:13}. Eq.~(\ref{WQori}) can be directly employed to analyze quantum thermodynamical cycles,
as an efficient way of assuring that no heat is exchanged in intermediate steps~\cite{Geva:92,Ting:06,Quan:07} or
to minimize quantum friction in a non-equilibrium setup~\cite{Esposito:10,Deffner:10,Plastina:14}.

Here, we theoretically and experimentally discuss thermodynamical adiabatic processes in real (open) quantum systems evolving under decoherence.
To this end, we address the problem from a general approach for adiabatic dynamics in decohering systems. In contrast with closed systems,
heat may be exchanged in the case of non-unitary evolution. In particular, we will establish a sufficient condition to ensure that
an adiabatic open-system dynamics (associated with Lindblad-Jordan decoupled eigenspaces) leads to an adiabatic thermodynamical process
(associated with no heat exchange). Moreover, for thermodynamically non-adiabatic processes, we evaluate the von Neumann 
entropy, discussing its relation with heat for arbitrary evolution time. Our results are then experimentally implemented by using a hyperfine energy-level quantum bit (qubit) of an Ytterbium $^{171}$Yb$^+$ trapped ion,
where reservoir engineering is performed to achieve a controllable adiabatic dynamics. Due to requirements of the usual definitions of heat and work,
the investigation of thermodynamic quantities in adiabatic dynamics is achieved with time-dependent decoherence effects.
To this end, we introduce an efficient control to a Gaussian noise with time-dependent amplitude, which is then used to simulate a dephasing
channel with a time-dependent decoherece rate $\gamma(t)$.

\section*{Results} 

\noindent\textbf{Work and heat in the adiabatic dynamics of open systems} 

\vspace{0.3cm}

We start by introducing heat and work in a general formalism for adiabaticity in open quantum systems, namely, the superoperator
formalism~\cite{Sarandy:05-1}. In this work, we will consider a discrete quantum systems ${\cal S}$ defined over a $d$-dimensional
Hilbert space. The system ${\cal S}$ interacts with its surrounding environment ${\cal A}$. The dynamics is assumed to be
described by a time-local master equation $\dot{\rho}(t)\!=\!\Lcal_{t} [\rho (t)]$, where $\rho(t)$ is the density operator associated with ${\cal S}$
and  $\Lcal_{t} [\bullet]$ is a time-dependent Liouville operator. The Liouville operator takes the form
$\Lcal_{t} [\rho (t)]\!=\!\Hcal_{t} [\rho (t)] + \Rcal_{t} [\rho (t)]$, where $\Hcal_{t}[\bullet]\!=\!(1/i\hbar) [H(t),\bullet]$ is the unitary
part of the dynamics and $\Rcal_t [\bullet]$ describes the decohering effects of ${\cal A}$ over ${\cal S}$.

In the superoperator formalism, the open-system dynamics can be provided from a Schrödinger-like equation
$\dket{\dot{\rho} (t)}\!=\!\Lmath(t) \dket{\rho (t)}$,
where $\Lmath(t)$ is termed
the Lindblad
superoperator and the density operator $\dket{\rho(t)}$ is represented by a $D^2$-dimensional vector (hence the double ket notation), whose components $\varrho_{k}(t)$ can be suitably expanded in terms of tensor products of the Pauli basis $\{\1,\sigma_{1},\sigma_{2},\sigma_{3}\}$~\cite{Sarandy:05-1}.
For instance, for the case of a single qubit ($D\!=\!2$), we have  $\rho(t)\!=\! \frac{1}{2}\sum_{k=0}^{3} \varrho_{k}(t)\sigma_{k}$ and
$\varrho_{k}(t)\!=\!\trs{\rho(t) \sigma_{k}}$,
with $\sigma_{k}$ denoting an element of the Pauli basis.
Moreover, $\Lmath(t)\!=\!\Hmath(t) + \Rmath(t)$, where $\Hmath(t)$ and $\Rmath(t)$ are $(D^2\times D^2)$-dimensional {super-matrices},
whose elements are $\Hmath_{ki}(t)\!=\!(1/D)\text{Tr}\{\sigma_{k}^{\dagger}\Hcal_{t}[\sigma_{i}]\}$ and $\Rmath_{ki}(t)\!=\!(1/D) \text{Tr}\{ \sigma_{k}^{\dagger} \Rcal [ \sigma_{i} ] \}$, respectively. The thermodynamic quantities defined in Eq.~\eqref{delQ} are then rewritten as
(see Methods section)
\begin{equation}
d W_{\mathrm{op}} = \frac{1}{D}\dinterpro{\dot{h}(t)}{\rho(t)} dt, \,\,\,\,
d Q_{\mathrm{op}} =  \frac{1}{D}\dinterpro{h(t)}{\Lmath (t)|\rho(t)}dt, \label{eq2}
\end{equation}
with the components $h_{k}(t)$ of $\dbra{h(t)}$ defined by $h_{k}(t)=\trs{H(t) \sigma_{k}}$.
In this notation, the inner product of vectors $|u\rangle\rangle$ and $|v\rangle\rangle$ associated with operators $u$ and $v$,
respectively, is defined as $\langle\langle u|v\rangle\rangle = (1/D) \textrm{Tr}(u^\dagger v)$.

Because $\Lmath (t)$ is non-Hermitian, it cannot always be diagonalized. Then, the definition of adiabaticity in this scenario is subtler than in the case of closed systems.
For open systems,  the adiabatic dynamics can be defined in terms of the Jordan decomposition of  $\Lmath (t)$~\cite{Sarandy:05-1}. More specifically,
adiabaticity is associated with a completely positive trace-preserving dynamics that can be decomposed into decoupled Lindblad-Jordan eigenspaces
associated with distinct non-crossing instantaneous eigenvalues $\lambda_{i}(t)$ of $\Lmath (t)$. We notice here that some care is required in order to find a basis for describing the density operator. The standard technique is to start from the instantaneous right and left eigenstates of $\Lmath (t)$, completing these eigensets in order to compose right $\{\dket{\Dcal^{(k_{i})}_{i}(t)}\}$
and left $\{\langle\langle{\cal{E}}^{(k_{i})}_{i}(t)|\}$ vector bases, where $\dket{\Dcal^{(k_{i})}_{i}(t)}$ and
$\langle\langle{\cal{E}}^{(k_{i})}_{i}(t)|$ are the $k_{i}$-th right and left vectors, respectively, associated with the eigenspace with eigenvalue $\lambda_{i}(t)$
in the Jordan decomposition of $\Lmath (t)$. These Jordan left and right bases can always be built such that they satisfy a bi-orthonormal relationship
$\langle\langle{\cal{E}}^{(\alpha)}_{i}(t)|{\cal{D}}^{(\beta)}_{j}(t)\rangle\rangle = \delta_{ij}\delta^{\alpha\beta}$.
Assuming an open-system adiabatic dynamics, we can analytically derive work, heat, and entropy variation.
Indeed, by taking the initial density operator as $\dket{\rho (0)} = \sum \nolimits _{i,k_{i}} c^{(k_{i})}_{i}\dket{\Dcal^{(k_{i})}_{i}(0)}$, we obtain that work and heat are provided by
\begin{align}
d W^{\mathrm{ad}} &= \frac{1}{D} \sum \nolimits _{i,k_{i}} c^{(k_{i})}_{i}e^{\int_{0}^{t} \tilde{\lambda}_{i,k_{i}}(t^{\prime})dt^{\prime}}
\dinterpro{\dot{h}(t)}{\Dcal^{(k_{i})}_{i}(t)} dt \label{dWadiab} \text{ , }  \\
d Q^{\mathrm{ad}} &= \frac{1}{D}\sum \nolimits _{i,k_{i}} c^{(k_{i})}_{i}e^{\int_{0}^{t} \tilde{\lambda}_{i,k_{i}}(t^{\prime})dt^{\prime}}\dbra{h(t)} \Lmath (t) \dket{\Dcal^{(k_{i})}_{i}(t)} dt \text{ , } \label{dQadiab}
\end{align}
with $d W^{\mathrm{ad}}$ ($d Q^{\mathrm{ad}}$) being identified
to the amount of work (heat) performed on/by the system.

The validity of Eqs.~\eqref{dWadiab} and~\eqref{dQadiab} is shown in the Methods section. As long as we are in the weak coupling regime and 
the system is driven by a time-local master equation, Eqs.~(\ref{dWadiab}) and~(\ref{dQadiab}) provide expressions for work and heat for the 
adiabatic decohering dynamics. Notice also that the adiabatic dynamics will require a slowly varying Liouville superoperator ${\Lmath (t)}$~\cite{Sarandy:05-1}. 
Starting from Eq.~(\ref{eq2}), we are allowed to evaluate the 
density operator $|\rho(t)\rangle \rangle$ through an 
arbitrary strategy. For instance, we could apply a piecewise deterministic process approach 
via Feynman-Vernon path integral for the corresponding propagator~\cite{Petruccione:Book}. 
Alternatively, we could implement a numerical simulation  
via a Monte Carlo wave function method~(see, e.g., Ref.~\cite{Plenio:98} and references therein). 
In all these cases, from Eqs.~(\ref{dWadiab}) and~(\ref{dQadiab}), we can obtain a sufficient condition for avoiding
heat exchange in a quantum mechanical adiabatic evolution. More specifically, if the initial state $\rho(0)$ of the system can be written as a superposition
of the eigenstate set $\{\dket{\Dcal^{(k_{i})}_{i}(0)}\}$ with eigenvalue $\lambda_{i}(t) = 0$, for every $t \in [0,\tau]$, the adiabatic dynamics implies in no heat exchange.
Therefore, we can establish that an adiabatic
	dynamics in quantum mechanics is not in general associated with an adiabatic process
	in quantum thermodynamics, with a sufficient condition for thermal adiabaticity being the evolution within an eigenstate set with vanishing eigenvalue of $\Lmath (t)$. 
	This condition is satisfied by a quantum system that adiabatically evolves under a steady state trajectory, 
	since such dynamics can be described by an eigenstate (or a superposition of eigenstates) of $\Lmath (t)$ with eigenvalue zero~\cite{Venuti:16}.  
	As an example, Ref.~\cite{Li:17} has considered the adiabatic evolution of 2D topological insulators, where the system evolves through its 
	steady state trajectory. For this system, the evolved state $\dket{\rho_{\text{ss}}(t)}$, associated with the steady state of the system 
	$\rho_{\text{ss}}(t)$, satisfies $\Lmath (t)\dket{\rho_{\text{ss}}(t)} = 0$, $\forall t$. This means that $\dket{\rho_{\text{ss}}(t)}$ is an 
	instantaneous eigenstate of $\Lmath (t)$ with eigenvalue $\lambda(t) = 0$.

\vspace{0.5cm}

\noindent\textbf{Thermal adiabaticity for a qubit adiabatic dynamics} 

\vspace{0.3cm}

As a further illustration, let us consider a two-level system initialized in a thermal equilibrium
state $\rho_{\text{th}}(0)$ for the Hamiltonian $H(0)$ at inverse temperature $\beta\!=\!1/k_{\text{B}}T$, where $k_{\text{B}}$ and $T$ are the
Boltzmann's constant and the absolute temperature, respectively. Let the system be governed by a Lindblad
equation, where the environment acts as a {dephasing} channel in the energy eigenstate basis
$\{\ket{E_{n}(t)}\}$ of $H(t)$. Thus, we describe the coupling between the system and its reservoir through
$\Rcal^{\text{dp}}_{t} [\bullet]\!=\!\gamma (t) [ \Gamma^{\text{dp}}(t) \bullet \Gamma^{\text{dp}}(t) - \bullet ]$,
where $\Gamma^{\text{dp}}(t)\!=\!\ket{E_{0}(t)}\bra{E_{0}(t)} - \ket{E_{1}(t)}\bra{E_{1}(t)}$.
In this case, the set of eigenvectors of $\Lmath (t)$
can be obtained from set of operators $P_{nm}(t)\!=\!\ket{E_{n}(t)}\bra{E_{m}(t)}$, where the components $\Dcal^{(i)}_{nm}(t)$ of $\dket{\Dcal_{nm}(t)}$
are given by $\Dcal^{(i)}_{nm}(t)\!=\!\trs{P_{nm}(t)\sigma_{i}}$. Moreover, the eigenvalue equation for $\Lmath (t)$ can be written as
$\Lmath (t) \dket{\Dcal_{nm}(t)}\!=\!\lambda_{nm}(t) \dket{\Dcal_{nm}(t)}$, where $\lambda_{nm}(t)\!=\!E_{n}(t)-E_{m}(t) - 2(1-\delta_{nm})\gamma(t)$.
In the superoperator formalism, the initial state $\rho_{\text{th}}(0)$ is written as
$\dket{\rho_{\text{th}}(0)}\!=\!\Zcal^{-1}(0) \sum\nolimits_{n} e^{-\beta E_{n}(0)}\dket{\Dcal_{nn}(0)}$, where $\Zcal(t)\!=\!\trs{e^{-\beta H(t)}}$ is the
partition function of the system. Therefore, since $\dket{\rho_{\text{th}}(0)}$ is given by a superposition of eigenvectors of $\Lmath (t)$ with
eigenvalue $\lambda_{nn}(t)\!=\!0$, we obtain from Eq.~(\ref{dQadiab}) that $d Q^{\mathrm{ad}}\!=\!0$. Therefore,
thermal adiabaticity is achieved for an arbitrary open-system adiabatic dynamics subject to dephasing in the energy eigenbasis.
Hence, any internal energy variation for this situation should be identified as {work}.

\vspace{0.5cm}

\noindent\textbf{Heat exchange for a qubit adiabatic dynamics}

\vspace{0.3cm}

In contrast, we can use a similar qubit system to find a process in which heat can be exchanged, i.e., $d Q^{\mathrm{ad}}\!\neq\!0$.
To this end, let us consider dephasing in the computational basis, with the coupling between the system and its reservoir through
$\Rcal^{\text{z}}_{t} [\bullet]\!=\!\gamma (t) \left[ \sigma_{z} \bullet \sigma_{z} - \bullet \right]$. In order to guarantee that any internal
energy variation is associated to heat exchange, we consider a constant Hamiltonian during the entire non-unitary evolution
(so that $d W^{\mathrm{ad}}\!=\!0$). Since
$\Rcal^{\text{z}}_{t} [\bullet]$ must not be written in the eigenbasis of the Hamiltonian, we assume a Hamiltonian $H_{\text{x}}\!=\!\hbar \omega \sigma_{x}$,
where the system is initialized in the typical initial state of a thermal machine, namely, the thermal state of the Hamiltonian $H_{\text{x}}$ at some arbitrary temperature
$\beta$. By letting the system undergo a non-unitary adiabatic dynamics under dephasing, the evolved state is~(see Methods section)
\begin{eqnarray}
\rho^{\text{ad}}(t) = \frac{1}{2} \left[\1 - e^{-2 \int_{0}^{t}\gamma(\xi)d\xi}\tanh(\beta \hbar \omega) \sigma_{x}\right] \text{ . } \label{rhox}
\end{eqnarray}

From Eq.~\eqref{dQadiab} we then compute the amount of exchanged heat during an infinitesimal time interval $dt$ as 
$
d Q^{\mathrm{ad}}(t) = 2 \hbar \tanh(\beta \hbar \omega) \omega \gamma(t) e^{-2 \int_{0}^{t}\gamma(\xi)d\xi} dt
$.
The negative argument in the exponential shows that the higher the mean-value of $\gamma(t)$ the faster the heat exchange ends~(see Methods section). Thus, if we define the amount of exchanged heat during the entire evolution as $\Delta Q(\tau_{\text{dec}})\!=\!\int_{0}^{\tau_{\text{dec}}} [dQ^{\mathrm{ad}}(t)/dt]dt$,
where $\tau_{\text{dec}}$ is the total evolution time of the nonunitary dynamics, we get
\begin{align}
\Delta Q(\tau_{\text{dec}}) = \hbar \omega \tanh(\beta \hbar \omega) \left( 1 - e^{-2 \bar{\gamma} \tau_{\text{dec}}}\right) \text{ , }
\label{Q}
\end{align}
{where $\bar{\gamma} = (1/\tau_{\text{dec}})\int_0^{\tau_{\text{dec}}} \gamma(\xi) d\xi$ is the average dephasing rate during $\tau_{\text{dec}}$.
Notice that $\Delta Q(\tau_{\text{dec}})\!>\!0$ for any value of $\bar{\gamma}$. Therefore, the dephasing channel considered
here works as an artificial thermal reservoir at inverse temperature $\tilde{\beta}\!=\!\beta_{\text{deph}}\!<\!\beta$, 
 with $\beta_{\text{deph}}\!=\!(1/\hbar \omega) \text{arctanh} [ e^{-2 \bar{\gamma} \tau_{\text{dec}}}\tanh(\beta \hbar \omega) ]$~(see Methods section).
 We can further compute the maximum exchanged
heat from Eq.~\eqref{Q} as a quantity independent of the environment parameters and given by $\Delta Q_{\text{max}}\!=\!\hbar \omega \tanh(\beta \hbar \omega)$.
It would be worth to highlight that, for quantum thermal machines weakly coupled to thermal reservoirs at different temperatures \cite{Alicki:79},
the maximum heat $\Delta Q_{\text{max}}$ is obtained with high-temperature hot reservoirs~\cite{Geva:92,Henrich:07,Lutz:16-Science}.

Despite we have provided a specific open-system adiabatic evolution, we can determine infinite classes
of system-environment interactions exhibiting the same amount of heat exchange $d Q$. In particular, there are infinite engineered environments 
that are able to extract a maximum heat amount $\Delta Q_{\text{max}}$. A detailed proof of this result can be found in Methods section.

\vspace{0.5cm}

\noindent\textbf{Experimental realization} 

\vspace{0.3cm}

We now discuss an experimental realization to test the thermodynamics of adiabatic processes in an open-system evolution. This is
implemented using the hyperfine energy levels of an Ytterbium ion $^{171}$Yb$^+$ confined by a six-needles Paul trap,
with a qubit encoded into the $^{2}S_{1/2}$ ground state,
$\ket{0}\!\equiv\!\ket{^{2}S_{1/2};\, F\!=\!0,m_{F}\!=\!0}$ and  $\ket{1}\!\equiv\!\,\ket{^{2}S_{1/2};\, F\!=\!1,m_{F}\!=\!0}$,
as shown in Fig. ({\color{blue}1a})~\cite{Olmschenk:07}. The qubit initialization is obtained from the standard Rabi Oscillation sequence~\cite{Olmschenk:07}, where we first implement the Doppler cooling for $1$~ms, after we apply a standard optical pumping process for $0.01$~ms to initialize the qubit into the $\ket{0}$ state, and then we use microwave to implement the desired dynamics. The target Hamiltonian $H_{\text{x}}$ can be realized using a resonant microwave with Rabi frequency adjusted to $\omega$. To this end, the channel 1 (CH1) waveform of a programmable two-channel arbitrary waveform generator (AWG) is used, which has been programmed to the angular
frequency $2\pi \times 200$~MHz. As depicted in Fig. ({\color{blue}1b}), to implement the dephasing channel we use the Gaussian noise
frequency modulation (FM) microwave technique, which has been developed in a recent previous work and shows high controllability~\cite{Hu:18}. Since we need to implement a time-dependent decohering quantum channel, we use the channel 2 (CH2) waveform as amplitude modulation (AM) source to achieve high control of the Gaussian noise amplitude, consequently, to optimally control of the dephasing rate $\gamma(t)$. The dephasing rates are calibrated by fitting the Rabi oscillation curve with exponential decay. Since the heat flux depend on the non-unitary process induced by the system-reservoir coupling, then by using a different kind of noise (other than the Gaussian form) we may obtain a different heat exchange behavior. See Methods section for a detailed description of the experimental setup, including the
implementation of the quantum channel and the quantum process tomography~(see Methods section).

As a further development, we analyze in Fig.~\ref{Graph1} the experimental results for the heat exchange $\Delta Q(\tau_{\text{dec}})$ as a function of $\tau_{\text{dec}}$,
where we have chosen $\gamma(t)\!=\!\gamma_{0}(1+t/\tau_{\text{dec}})$, where $\tau_{\text{dec}}$ is experimentally controlled through the time interval associated to the action of our decohering quantum channel. 
The solid curves in Fig.~\ref{Graph1} are computed from Eq.~\eqref{Q}, while the experimental points are computed through the variation of internal energy as $\Delta Q(\tau_{\text{dec}})\!=\!U_{\text{fin}} - U_{\text{ini}}$, where $U_{\text{fin(ini)}}\!=\!\trs{\rho_{\text{fin(ini)}}H(\tau)}$. The computation of $U_{\text{fin(ini)}}$ is directly obtained from quantum state tomography of $\rho_{\text{fin(ini)}}$ for each value of $\tau_{\text{dec}}$. Although the maximum exchanged heat is independent of $\gamma_0$, the initial dephasing rate $\gamma_0$ affects the {power} for which the system exchanges heat with the reservoir for a given evolution time $\tau_{\text{dec}}$ (See Methods section). Thus, since we have an adiabatic path in open system~(see Methods section), the curves in Fig.~\ref{Graph1} represent the heat exchanged during the adiabatic dynamics. It is worth highlighting here that we can have different noise sources in the trapped ion system in addition to dephasing. However, the coherence timescale of the Ytterbium hyperfine qubit is around 200~ms~\cite{Hu:18}. Therefore, 
it is much larger than the timescale of the experimental implementation. Indeed, the dephasing rates implemented 
in our realization are simulated by the experimental setup.

As previously mentioned, since the Hamiltonian is time-independent, any internal energy variation is identified as heat. In order to provide a more detailed view of this heat exchange, 
we analyze the von Neumann entropy $S(\rho) = - \textrm{tr} \,(\rho \log \rho)$ during the evolution. To this end, by adopting the superoperator formalism as before, the entropy variation for an infinitesimal time interval $dt$ reads 
$dS\!=\!- (1/D)\dbra{\rho_{\log}(t)}\Lmath (t) \dket{\rho(t)}$, where $\dbra{\rho_{\log}(t)}$ is a supervector with components given by $\varrho^{\log}_{n}(t)\!=\! \tr{\sigma_{n} \log \rho(t)}$~(see Methods section). Thus, 
for an adiabatic evolution in an open system we find that~(see Methods section)
\begin{align}
d S &=-\frac{1}{D}\sum \nolimits _{i,k_{i}} c^{(k_{i})}_{i} e^{\int_{0}^{t} \tilde{\lambda}_{i,k_{i}}(t^{\prime})dt^{\prime}}
\Gamma_{i,k_{i}}(t) \text{ , } \label{dEntropyPro}
\end{align}
where $\Gamma_{i,k_{i}}(t)\!=\! \dinterpro{\rho^{\text{ad}}_{\log}(t)}{\Dcal^{(k_{i}-1)}_{i}(t)} + \lambda_{i}(t) \dinterpro{\rho^{\text{ad}}_{\log}(t)}{\Dcal^{(k_{i})}_{i}(t)}$, with
$\dbra{\rho^{\text{ad}}_{\log}(t)}$ defined here as a supervector with components $\varrho^{\text{ad}}_{\log}(t)\!=\!\trs{\sigma_{n} \log \rho^{\text{ad}}_{\log}(t)}$. 
For the adiabatic dynamics considered in Fig.~\ref{Graph1} the infinitesimal von Neumann entropy variation $d S$ in interval $dt$ is given by
\begin{align}
dS(t) = 2 g(t) \gamma(t) \text{arctanh} [g(t)] dt \text{ , }
\end{align}
where we define $g(t)\!=\!e^{-2 \int_{0}^{t}\gamma(\xi)d\xi}\tanh(\beta \hbar \omega)$. Notice that the relation between heat and entropy can be obtained by rewriting the exchanged heat $d Q$ in the interval $dt$ as 
$d Q^{\mathrm{ad}}(t) = 2 \hbar \omega \gamma(t) g(t)  dt$. In conclusion, the energy variation can indeed be identified as heat exchanged along the adiabatic dynamics. Indeed, by computing the thermodynamic relation between $dS(t)$ and $dQ^{\text{ad}}(t)$ we get $dS(t) = \beta_{\text{deph}} dQ^{\text{ad}}(t)$, where $\beta_{\text{deph}}$ is the  inverse temperature of the simulated thermal bath.

\section*{Discussion} 

From a general approach for adiabaticity in open quantum systems driven by time-local master equations, 
we provided a relationship between adiabaticity in quantum mechanics and
in quantum thermodynamics in the weak coupling regime between system and reservoir. In particular, we derived a sufficient condition for which the adiabatic dynamics in open quantum systems leads to adiabatic
processes in thermodynamics. By using a particular example of a single qubit undergoing an open-system adiabatic evolution path, we have illustrated the existence of both
adiabatic and diabatic regimes in quantum thermodynamics, computing the associated heat fluxes in the processes. As a further result, we also proved the existence of an infinite family of decohering systems exhibiting the same maximum heat exchange.
From the experimental side, we have realized adiabatic open-system evolutions using an Ytterbium trapped ion, with its hyperfine energy level encoding
a qubit (work substance).
In turn, we have experimentally shown that heat exchange can be directly provided along the adiabatic path in terms of the decoherence rates as a function of the total evolution time. In particular, the relationship between heat and entropy is naturally derived in terms of a simulated thermal bath.
Our implementation exhibits high controllability, opening perspectives for analyzing thermal machines (or refrigerators) in
open quantum systems under adiabatic evolutions.
Moreover, a further point to be explored is the speed up of the adiabatic path through the transitionless quantum driving (TQD) 
	method for open systems~\cite{Vacanti:14}. Indeed, TQD can be 
	incorporated in the formalism for adiabatic thermodynamics we introduced in this work. The starting point is the generalization of Eqs.~(\ref{dWadiab}) and~(\ref{dQadiab}) 
	through the introduction of the superadiabatic Lindbladian superoperator ${\Lmath}_{\textrm{TQD}}(t)$ governing the open system evolution~\cite{Vacanti:14}. 
	Notice that ${\Lmath}_{\textrm{TQD}}(t)$ will include counter-diabatic contributions generally obtained by reservoir engineering. 
	Suppression of heat may be possibly obtained by constraining the evolution inside the Jordan block of ${\Lmath}_{\textrm{TQD}}(t)$ 
	with vanishing eigenvalue. Naturally, the requirements of weak coupling and time-local 
	master equations are still to be kept. The associated effects of the engineered reservoirs on the thermal efficiencies and TQD 
	dynamics are left for future research.

\section{Methods}


\noindent\textbf{Thermodynamics in the superoperator formalism} 

\vspace{0.3cm}

Let us consider the heat exchange as
\begin{equation}
dQ_{\text{op}} = \trs{\dot{\rho} (t) H(t)} dt = \trs{\Lcal[\rho (t)] H(t)} dt \label{dQ1open} \text{ .}
\end{equation}
where we have used the equation $\dot{\rho} (t) = \Lcal[\rho (t)]$. To derive the corresponding expression in the superoperator formalism we first define the basis of operators given by $\{\sigma_{i}\}$, $i=0,\cdots,D^2-1$, where $\trs{\sigma^{\dagger}_{i}\sigma_{j}} = D\delta_{ij}$. In this basis, we can write $\rho (t)$ and $H(t)$ generically as
\begin{align}
H(t) = \frac{1}{D}\sum_{n=0}^{D^2-1} h_{n}(t)\sigma ^{\dagger}_{n} \text{ \, and \, } \rho(t) = \frac{1}{D}\sum_{n=0}^{D^2-1} \varrho_{n}(t)\sigma_{n} \text{ ,} \label{HroSuper}
\end{align}
where we have $h_{n}(t) = \trs{H(t)\sigma_{n}}$ and $\varrho_{n}(t) = \trs{\rho(t)\sigma^{\dagger}_{n}}$. Then, we get
\begin{align}
dQ_{\text{op}} &= \frac{1}{D^2}\left( \sum _{n,m=0}^{D^2-1}  \trs{\Lcal[\varrho_{n}(t)\sigma_{n}] h_{m}(t)\sigma ^{\dagger}_{m}} \right) dt \nonumber \\&= \frac{1}{D^2}\left( \sum _{n,m=0}^{D^2-1} \varrho_{n}(t) h_{m}(t) \trs{\Lcal[\sigma_{n}] \sigma ^{\dagger}_{m}} \right) dt \label{dQ2open} \text{ .}
\end{align}
Now, we use the definition of the matrix elements of the superoperator $\Lmath(t)$, associated with $\Lcal[\bullet]$, which reads $\Lmath_{mn} = (1/D)\trs{\sigma ^{\dagger}_{m}\Lcal[\sigma_{n}] }$, so that we write
\begin{align}
dQ_{\text{op}} &= \frac{1}{D}\left( \sum _{n,m=0}^{D^2-1} h_{m}(t) \Lmath_{mn} \varrho_{n}(t) \right) dt \label{dQ3open} \text{ .}
\end{align}
In conclusion, by defining the vector elements
\begin{align}
\dbra{h(t)} &= \begin{bmatrix}
h_{0}(t) & h_{1}(t) & \cdots & h_{D^2-1}(t)
\end{bmatrix}^{\text{t}} \text{ , } \label{BraH}\\
\dket{\rho(t)} &= \begin{bmatrix}
\varrho_{0}(t) & \varrho_{1}(t) & \cdots & \varrho_{D^2-1}(t)
\end{bmatrix}\text{ , } \label{KetRho}
\end{align}
we can rewrite Eq.~\eqref{dQ3open}, yielding
\begin{equation}
dQ_{\text{op}} = \frac{1}{D}\dinterpro{h(t)}{\Lmath (t)|\rho(t)}dt \text{ . } \label{QAp}
\end{equation}
Equivalently,
\begin{equation}
dW_{\text{op}} = \trs{\rho (t) \dot{H}(t)} dt \label{dW1open} \text{ , }
\end{equation}
where we have used Eq.~\eqref{HroSuper} to write $\dot{H}(t) = (1/D)\sum_{n=0}^{D^2-1} \dot{h}_{n}(t)\sigma ^{\dagger}_{n}$ and, consequently,
\begin{equation}
dW_{\text{op}} =  \frac{1}{D}\sum_{n=0}^{D^2-1}\dot{h}_{n}(t)\trs{\rho (t) \sigma ^{\dagger}_{n}} dt \text{ , }
\end{equation}
so that we use the definition of the coefficients $\varrho_{n}(t)$ to get
\begin{equation}
dW_{\text{op}} =  \frac{1}{D}\sum_{n=0}^{D^2-1}\dot{h}_{n}(t)\varrho_{n}(t) dt \text{ . } \label{WAp}
\end{equation}
By using Eqs.~\eqref{BraH} and \eqref{KetRho} into Eq.~(\ref{WAp}), we conclude that
\begin{equation}
d W_{\mathrm{op}} = \frac{1}{D}\dinterpro{\dot{h}(t)}{\rho(t)} dt \text{ . } \label{WAp2}
\end{equation}

In thermodynamics, heat exchange is accompanied of an entropy variation. Then, in order to provide a complete thermodynamic study from this formalism, we now compute the instantaneous variation of the  
von Neumann entropy $S(t) = - \trs{\rho(t) \log [\rho(t)]}$, which reads
\begin{align}
\dot{S} (t) = - \frac{d}{dt}\left[ \trs{\rho(t) \log \rho(t)} \right] = - \trs{ \dot{\rho} (t)\log \rho(t) } - \trs{ \dot{\rho} (t)} \text{ . }
\end{align}
By using that $\trs{\rho(t)} = 1$, we get $\trs{ \dot{\rho} (t)} = 0 $. Therefore
\begin{align}
\dot{S} (t) = - \trs{ \dot{\rho} (t)\log \rho(t) } = - \trs{ \Lcal_{t} [\rho(t)] \log \rho(t) } \text{ , }
\end{align}
where we also used that $ \dot{\rho} (t) = \Lcal_{t} [\rho(t)]$. Now, let us to write
\begin{align}
\log \rho(t) &= \frac{1}{D}\sum_{n=0}^{D^2-1} \varrho_{n}^{\log}(t)\sigma_{n}^{\dagger} \text{ , } 
\end{align}
so that we can define the vectors $\dbra{\rho_{\log}(t)}$ associated to $\log \rho(t)$ with components $\varrho_{n}^{\log}(t)$ obtained as $\varrho_{n}^{\log}(t) = \trs{\sigma_{n}\log \rho(t)}$. Thus, we get
\begin{align}
\dot{S} (t) = -\frac{1}{D^2}\sum_{m=0}^{D^2-1}\sum_{n=0}^{D^2-1}\varrho_{m}(t) \varrho_{n}^{\log}(t)\trs{ \Lcal_{t} [\sigma_{m}]\sigma_{n}^{\dagger} } \text{ , }
\end{align}
In the superoperator formalism, we then have 
\begin{align}
\dot{S} (t) = - \frac{1}{D}\dbra{\rho_{\log}(t)}\Lmath (t) \dket{\rho(t)}  \text{ . }  \label{VarSAp}
\end{align}
Alternatively, it is possible to get a similar result for the entropy variation in an interval $\Delta t = t - t_{0}$ as
\[
\Delta S(t,t_{0}) = S(t) - S(t_{0}) = \tr{\rho(t_{0}) \log \rho(t_{0}) - \rho(t) \log \rho(t)} \text{ , }
\]
where we can use Eq.~\eqref{HroSuper} to write
\begin{align}
\Delta S(t,t_{0}) &= \frac{1}{D}\sum_{n=0}^{D^2-1} \varrho_{n}(t_{0}) \tr{\sigma_{n} \log \rho(t_{0})} \nonumber \\&- \frac{1}{D}\sum_{n=0}^{D^2-1} \varrho_{n}(t) \tr{\sigma_{n} \log \rho(t)} \text{ , }
\end{align}
so that we can identify $\varrho^{\log}_{n}(t) = \tr{\sigma_{n} \log \rho(t)}$ and we finally write
\begin{align}
\Delta S(t,t_{0}) &= \frac{1}{D}\sum_{n=0}^{D^2-1} \varrho_{n}(t_{0})\varrho^{\log}_{n}(t_{0}) - \frac{1}{D}\sum_{n=0}^{D^2-1} \varrho_{n}(t) \varrho^{\log}_{n}(t) \nonumber \\&= \frac{1}{D} \left[ \dinterpro{\rho_{\log}(t)}{\rho(t)} - \dinterpro{\rho_{\log}(t_{0})}{\rho(t_{0})}\right] \text{ . }
\end{align}

\vspace{0.5cm}

\noindent\textbf{Adiabatic quantum thermodynamics} 

\vspace{0.3cm}

 Let us start by briefly reviewing the adiabatic dynamics in the context of open systems. To this end, let us consider the local master equation (in the superoperator formalism)
\begin{align}
\dot{\rho} = \Lcal[\rho (t)] \text{ , } \label{MasterEq}
\end{align}
which describes a general time-local physical process in open systems. The dynamical generator $\Lcal[\bullet]$ is requested to be a linear operation, namely,
\begin{align}
\Lcal[\alpha_{1}\rho_{1}(t) + \alpha_{2}\rho_{2}(t)] = \alpha_{1}\Lcal[\rho_{1}(t)] + \alpha_{2}\Lcal[\rho_{2}(t)] \text{ , }
\end{align}
for any complex numbers $\alpha_{1,2}$ and matrices $\rho_{1,2}(t)$, with $\alpha_{1}+\alpha_{2} =1$, because we need to satisfy $\tr{ \alpha_{1}\rho_{1}(t) + \alpha_{2}\rho_{2}(t)}=1$. 
Thus, by using this property of the operator $\Lcal[\bullet]$, it is possible to rewrite Eq.~\eqref{MasterEq} as~\cite{Sarandy:05-1}
\begin{align}
\dket{\dot{\rho}(t)} = \Lmath (t) \dket{\rho(t)} \text{ , } \label{SuperMasterEq}
\end{align}
where $\Lmath (t)$ and $\dket{\rho(t)}$ have been already previously defined. In general, due to the non-Hermiticity of $\Lmath (t)$, there are situations in which $\Lmath (t)$ cannot be diagonalized, but it is always 
possible to write a block-diagonal form for $\Lmath (t)$ via the Jordan block diagonalization approach~\cite{Horn:Book}. Hence, it is possible to 
define a set of right and left {quasi}-eigenstates of $\Lmath (t)$, respectively, as
\begin{subequations}
	\label{EqEqEigenStateL}
	\begin{align}
	\Lmath(t)\dket{\Dcal_{\alpha}^{n_{\alpha}}(t)} &= \dket{\Dcal_{n}^{(n_{\alpha}-1)}(t)} + \lambda_{\alpha}(t)\dket{\Dcal_{\alpha}^{n_{\alpha}}(t)} \text{ , } \\
	\dbra{\Ecal_{\alpha}^{n_{\alpha}}(t)}\Lmath(t) &= \dbra{\Ecal_{n}^{(n_{\alpha}+1)}(t)} + \dbra{\Ecal_{\alpha}^{n_{\alpha}}(t)}\lambda_{\alpha}(t) \text{ . }
	\end{align}
\end{subequations}
From the above equations, we can write the Jordan form of $\Lmath(t)$ as
\begin{align}
\Lmath_{\text{J}}(t) = \text{diag}\begin{bmatrix}J_{1}(t) & J_{2}(t) & \cdots & J_{N}(t)\end{bmatrix} \text{, }
\end{align}
where $N$ is the number of distinct eigenvalues $\lambda_{\alpha}(t)$ and each block $J_{\alpha}(t)$ is given by
\begin{align}
J_{\alpha}(t) = 
\begin{bmatrix}
\lambda_{\alpha}(t) & 1   & 0        & \cdots & 0 \\
0 &\lambda_{\alpha}(t) & 1 & \cdots & 0 \\
\vdots & \ddots & \ddots & \ddots & \vdots \\
0 & \cdots & 0 & \lambda_{\alpha}(t) & 1  \\
0 & \cdots & \cdots & 0 & \lambda_{\alpha}(t)  
\end{bmatrix} \text{ . } \label{JDiago}
\end{align}
In the adiabatic dynamics of closed systems, the decoupled evolution of the set of eigenvectors $\ket{E_{n}^{k_{n}}(t)}$ of the Hamiltonian associated with an eigenvalue $E_{n}(t)$, where $k_{n}$ denotes individual eigenstates, characterizes what we call 
{Schr\"odinger-preserving eigenbasis}. In an analogous way, the set of right and left {quasi}-eigenstates of $\Lmath (t)$ associated with the Jordan block $J_{\alpha}(t)$ characterizes the {Jordan-preserving left and right bases}. Here, we will restrict our analysis to a particular case where each block $J_{\alpha}(t)$ is one-dimensional, so that the set of {quasi}-eigenstates given in Eq.~\eqref{EqEqEigenStateL} becomes a genuine eigenstate equation given by
\begin{subequations}
	\label{EqEqEigenOneL}
	\begin{align}
	\Lmath(t)\dket{\Dcal_{\alpha}(t)} &= \lambda_{\alpha}(t)\dket{\Dcal_{\alpha}(t)} \text{ , } \\
	\dbra{\Dcal_{\alpha}(t)}\Lmath(t) &= \dbra{\Ecal_{\alpha}(t)}\lambda_{\alpha}(t) \text{ . }
	\end{align}
\end{subequations}
In this case, we can expand the matrix density $\dket{\rho(t)}$ in basis $\dket{\Dcal_{\alpha}(t)}$ as
\begin{align}
\dket{\rho(t)} = \sum_{\alpha=1}^{N} r_{\alpha}(t)\dket{\Dcal_{\alpha}(t)} \text{ , }
\end{align}
with $r_{\beta}(t)$ being parameters to be determined. By using the Eq.~\eqref{SuperMasterEq}, one gets the dynamical equation for each $r_{\beta}(t)$ as
\begin{align}
\dot{r}_{\beta}(t) & = \lambda_{\beta}(t)r_{\beta}(t) - r_{\beta}(t) \dinterpro{\Ecal_{\beta}(t)}{\dot{\Dcal}_{\beta}(t)} \nonumber \\&- \sum_{\alpha\neq \beta}^{N} r_{\alpha}(t) \dinterpro{\Ecal_{\beta}(t)}{\dot{\Dcal}_{\alpha}(t)} \text{ . } \label{EqrdotOne}
\end{align}
Now, we can define a new parameter $p_{\beta}(t)$ as
\begin{align}
r_{\beta}(t) = p_{\beta}(t) e^{\int_{t_{0}}^{t} \lambda_{\beta}(\xi)d\xi} \text{ , } \label{pDef}
\end{align}
so that one finds an equation for $p_{\beta}(t)$ given by
\begin{align}
\dot{p}_{\beta}(t) &= - \sum_{\alpha\neq \beta}^{N} p_{\alpha}(t) e^{\int_{t_{0}}^{t} \lambda_{\alpha}(\xi) - \lambda_{\beta}(\xi) d\xi}\dinterpro{\Ecal_{\beta}(t)}{\dot{\Dcal}_{\alpha}(t)} \nonumber \\&- p_{\beta}(t) \dinterpro{\Ecal_{\beta}(t)}{\dot{\Dcal}_{\beta}(t)}\text{ , } \label{EqpdotOne}
\end{align}
with the first term in right-hand-side being the responsible for coupling distinct Jordan-Lindblad eigenspaces during the evolution. If we are able to apply some strategy to minimize the effects of such a term in the above equation, we can approximate the dynamics to
\begin{align}
\dot{p}_{\beta}(t) \approx - p_{\beta}(t) \dinterpro{\Ecal_{\beta}(t)}{\dot{\Dcal}_{\beta}(t)} \text{ . }
\end{align}
Then, the adiabatic solution $r_{\beta}(t)$ for the dynamics can be immediately obtained from Eq.~\eqref{pDef}, which reads
\begin{align}
r_{\beta}(t) = r_{\beta}(t_{0}) e^{\int_{t_{0}}^{t} \lambda_{\beta}(\xi)d\xi}
e^{- \int_{t_{0}}^{t} \dinterpro{\Ecal_{\beta}(\xi)}{\dot{\Dcal}_{\beta}(\xi)}d\xi} \text{ . } \label{Eqr1BJdec}
\end{align}
where we already used $p_{\beta}(t_{0}) = r_{\beta}(t_{0})$. In conclusion, if the system undergoes an adiabatic dynamics along a non-unitary process, the evolved state can be written as
\begin{align}
\dket{\rho^{\text{ad}}(t)} = \sum _{\alpha=1}^{N} r_{\alpha}(t_{0}) e^{\int_{t_{0}}^{t} \tilde{\lambda}_{\alpha}(\xi)d\xi}\dket{\Dcal_{\alpha}(t)} \label{EqAdEvol1D} \text{ , }
\end{align}
with $\tilde{\lambda}_{\alpha}(t)=\lambda_{\alpha}(t)-\dinterpro{\Ecal_{\alpha}(t)}{\dot{\Dcal}_{\alpha}(t)}$ being the generalized adiabatic phase accompanying the dynamics of the $n$-th eigenvector. 
The same mathematical procedure can be applied for multi-dimensional blocks~\cite{Sarandy:05-1}.
In this scenario, let $\dket{\rho (0)} = \sum \nolimits _{i,k_{i}} c^{(k_{i})}_{i}\dket{\Dcal^{(k_{i})}_{i}(0)}$ be the initial state of the system associated with the initial matrix density $\rho(0)$. 
By considering a general adiabatic evolution, the state at a later time $t$ will be given by~\cite{Sarandy:05-1}
\begin{align}
\dket{\rho^{\text{ad}}(t)} = \sum \nolimits _{i,k_{i}} c^{(k_{i})}_{i} e^{\int_{0}^{t} \tilde{\lambda}_{i,k_{i}}(t^{\prime})dt^{\prime}}\dket{\Dcal^{(k_{i})}_{i}(t)} \label{AdEvolvedAp}
\end{align}
with $\tilde{\lambda}_{i,k_{i}}(t)=\lambda_{i}(t)-\dinterpro{\Ecal^{(k_{i})}_{i}(t)}{\dot{\Dcal}^{(k_{i})}_{i}(t)}$, where $\{\dbra{\Ecal^{(k_{i})}_{i}(t)}\}$ and $\{\dket{\Dcal^{(k_{i})}_{i}(t)}\}$ denote the instantaneous Jordan-preserving left and right bases of $\Lmath (t)$, respectively~\cite{Sarandy:05-1}.
Therefore, from Eq.~\eqref{eq2}, we can write the work $d W_{\mathrm{op}}$  for an adiabatic dynamics as
\begin{equation}
d W_{\mathrm{op}} = \frac{1}{D}\sum \nolimits _{i,k_{i}} c^{(k_{i})}_{i} e^{\int_{0}^{t} \tilde{\lambda}_{i,k_{i}}(t^{\prime})dt^{\prime}}\dinterpro{\dot{h}(t)}{\Dcal^{(k_{i})}_{i}(t)} dt \text{ . }
\end{equation}
On the other hand, when no work is realized, we can obtain the heat $dQ_{\text{op}}$ for an adiabatic dynamics as
\begin{equation}
dQ^{\text{ad}} = \frac{1}{D}\sum \nolimits _{i,k_{i}} c^{(k_{i})}_{i} e^{\int_{0}^{t} \tilde{\lambda}_{i,k_{i}}(t^{\prime})dt^{\prime}}\dinterpro{h(t)}{\Lmath (t)|\Dcal^{(k_{i})}_{i}(t)}dt \text{ , } \label{QApAd}
\end{equation}
so that $dQ^{\text{ad}}$ represents the exchanged heat if no work is performed during such dynamics.
Moreover, from Eq.~\eqref{VarSAp}, we can write the von Neumann entropy variation as
\begin{align}
\dot{S} (t) &= - \frac{1}{D}\dbra{\rho^{\text{ad}}_{\log}(t)}\Lmath (t) \dket{\rho^{\text{ad}}(t)} \nonumber \\&= - \frac{1}{D}\sum \nolimits _{i,k_{i}} c^{(k_{i})}_{i} e^{\int_{0}^{t} \tilde{\lambda}_{i,k_{i}}(t^{\prime})dt^{\prime}}\dbra{\rho^{\text{ad}}_{\log}(t)}\Lmath (t) \dket{\Dcal^{(k_{i})}_{i}(t)} \nonumber \text{ , }
\end{align}
so that we can use the Eq.~\eqref{EqEqEigenStateL} to write
\begin{align}
\dot{S} (t) =\frac{1}{D}\sum \nolimits _{i,k_{i}} c^{(k_{i})}_{i} e^{\int_{0}^{t} \tilde{\lambda}_{i,k_{i}}(t^{\prime})dt^{\prime}}
\Gamma_{i,k_{i}}(t) \text{ , } \label{EntropyProAp}
\end{align}
where $\Gamma_{i,k_{i}}(t) = \dinterpro{\rho^{\text{ad}}_{\log}(t)}{\Dcal^{(k_{i}-1)}_{i}(t)} + \lambda_{i}(t) \dinterpro{\rho^{\text{ad}}_{\log}(t)}{\Dcal^{(k_{i})}_{i}(t)}$, with
$\dbra{\rho^{\text{ad}}_{\log}(t)}$ standing for the adiabatic evolved state associated with $\dbra{\rho_{\log}(t)}$.

\vspace{0.5cm}

\noindent\textbf{Heat in adiabatic quantum processes} 

\vspace{0.3cm}

We will discuss how to determine infinite classes of systems exhibiting the same amount of heat exchange $d Q$. This is provided in Theorem 1 below.

{Theorem 1 } ---
Let ${\cal S}$ be an open quantum system governed by a time-local master equation in the form
$\dot{\rho}(t) = \Hcal [\rho (t)] + \Rcal_{t}  [\rho (t)]$, where $\Hcal[\bullet] = (1/i\hbar) [H,\bullet]$ and
$\Rcal_{t}[\bullet]=\sum \nolimits _{n} \gamma_{n} (t) [ \Gamma_{n}(t) \bullet \Gamma^{\dagger}_{n}(t) - (1/2)\{  \Gamma^{\dagger}_{n}(t) \Gamma_{n}(t),\bullet\} ]$.
The Hamiltonian $H$ is taken as a constant operator so that no work is realized by/on the system.
Assume that the heat exchange between ${\cal S}$ and its reservoir during the quantum evolution is given by $d Q$.
Then, any unitarily related adiabatic dynamics driven by $\dot{\rho}^{\prime}(t) = \Hcal^{\prime} [\rho^{\prime} (t)] + \Rcal^{\prime}_{t}  [\rho^{\prime} (t)]$,
where $\dot{\rho}^\prime (t) = U \dot{\rho}(t) U^\dagger$, $\Hcal^{\prime}[\bullet] = U\Hcal[\bullet]U^{\dagger}$ and $\Rcal^{\prime}_{t}  [ \bullet] = U \Rcal_{t} [ \bullet]U^{\dagger}$, for some constant unitary $U$,
implies in an equivalent heat exchange $d Q^\prime = d Q$. $\square$

 {Proof } ---
Let us consider that $\rho(t)$ is solution of
\begin{align}
\dot{\rho}(t) &= \Hcal [\rho (t)] + \Rcal_{t}  [\rho (t)] \text{ , }
\end{align}
so, by multiplying both sides of the above equation by $U$ (on the left-hand-side) and $U^{\dagger}$ (on the right-hand-side), we get
\begin{align}
U\dot{\rho}(t)U^{\dagger} &= U\Hcal [\rho (t)]U^{\dagger} + U\Rcal_{t}  [\rho (t)]U^{\dagger} \nonumber \\
&= \frac{1}{i\hbar}U[H,\rho (t)]U^{\dagger} + \sum \nolimits _{n} \gamma_{n} (t) U\Gamma_{n}(t)\rho (t) \Gamma^{\dagger}_{n}(t)U^{\dagger} \nonumber \\
&- \frac{1}{2}\sum \nolimits _{n} \gamma_{n} (t)U\{  \Gamma^{\dagger}_{n}(t) \Gamma_{n}(t),\rho (t)\} ]U^{\dagger} \text{ , }
\end{align}
thus, by using the relations $[UAU^{\dagger},UBU^{\dagger}] = U[A,B]U^{\dagger}$ and $\{UAU^{\dagger},UBU^{\dagger}\} = U\{A,B\}U^{\dagger}$, we find
\begin{align}
\dot{\rho}^{\prime}(t) &= \frac{1}{i\hbar}[UHU^{\dagger},\rho^{\prime} (t)] + \sum \nolimits _{n} \gamma_{n} (t) \Gamma^{\prime}_{n}(t)\rho^{\prime} (t) \Gamma^{\prime\dagger}_{n}(t) \nonumber \\
&- \frac{1}{2}\sum \nolimits _{n} \gamma_{n} (t)\{  \Gamma^{\prime\dagger}_{n}(t) \Gamma^{\prime}_{n}(t),\rho^{\prime} (t)\}  \text{ , }
\end{align}
where $\Gamma^{\prime} (t) = U\Gamma_{n}(t)U^{\dagger}$. In conclusion, we get that $\rho^{\prime}(t) = U\rho (t)U^{\dagger}$ is a solution of
\begin{align}
\dot{\rho}^{\prime}(t) &= \Hcal^{\prime} [\rho^{\prime} (t)] + \Rcal^{\prime}_{t}  [\rho^{\prime} (t)] \text{ , }
\end{align}
where
\begin{align}
\Hcal^{\prime}[\bullet] &= \frac{1}{i\hbar}[UHU^{\dagger},\bullet] = U\Hcal[\bullet]U^{\dagger}, \\
\Rcal_{t}^{\prime}[\bullet] &= \sum \nolimits _{n} \gamma_{n} (t) [\Gamma^{\prime}_{n}(t)\rho^{\prime} (t) \Gamma^{\prime\dagger}_{n}(t) - \frac{1}{2}\{\Gamma^{\prime\dagger}_{n}(t) \Gamma^{\prime}_{n}(t),\rho^{\prime} (t)\} ] \nonumber \\
&= U\Rcal_{t}[\bullet]U^{\dagger} \text{ . }
\end{align}
Now, by taking into account that the Hamiltonian $H$ is a constant operator, we have that no work is realized by/on the system.
Then, by computing the amount of heat extracted from the system in the prime dynamics during an interval $t \in [0,\tau]$, we obtain
\begin{align}
\Delta Q^{\prime} = \trs{H^{\prime}\rho^{\prime}(\tau)} - \trs{H^{\prime}\rho^{\prime}(0)},
\end{align}
where, by definition, we can use $\rho^{\prime}(t) = U\rho(t)U^{\dagger}$, $ \forall t \in [0,\tau]$. Hence
\begin{align}
\Delta Q^{\prime} &= \trs{H^{\prime}U\rho(\tau)U^{\dagger}} - \trs{H^{\prime}U\rho(0)U^{\dagger}} \nonumber \\
&= \trs{U^{\dagger}H^{\prime}U\rho(\tau)} - \trs{U^{\dagger}H^{\prime}U\rho(0)} = \Delta Q
\end{align}
where we have used the cyclical property of the trace and that $\Delta Q = \trs{H\rho(\tau)} - \trs{H\rho(0)}$. $\blacksquare$

As an example of application of the above theorem, let us consider a system-reservoir interaction governed by $\Rcal^{\text{x}}_{t} [\bullet] = \gamma (t) \left[ \sigma_{x} \bullet \sigma_{x} - \bullet \right]$ (bit-flip channel). We can then show that the results previously obtained for dephasing can be reproduced if the quantum system is initially prepared in thermal state of $H_{\text{y}}^{0}=\omega\sigma_{y}$. Such a result is clear if we choose $U = R_{\text{x}}(\pi/2)R_{\text{z}}(\pi/2)$. Then, it follows that $\Rcal^{\text{x}}_{t}  [ \bullet] = U\Rcal^{\text{z}}_{t}[ \bullet]U^{\dagger}$ and $\Hcal^{\prime}[\bullet] = U\Hcal[\bullet]U^{\dagger}$, where $R_{\text{z(x)}}(\theta)$ are rotation matrices with angle $\theta$ around $z(x)$-axes for the case of a single qubit. Thus, the above theorem assures that the maximum exchanged heat will be $\Delta Q_{\text{max}} = \hbar \tilde{\omega} \tanh[\beta \hbar \omega]$.


Let us discuss now the adiabatic dynamics under dephasing and heat exchange.  
Consider the Hamiltonian $H_{\text{x}} = \hbar \omega \sigma_{x}$, where the system is initialized in the thermal of $H_{\text{x}}$ at inverse temperature
$\beta$. In this case, the initial state can be written as
\begin{align}
\rho(0) = \frac{1}{2} \left( \1 + \tanh[\beta \hbar \omega] \sigma_{x} \right) \text{ . }
\end{align}
If we rewrite the above state in superoperator formalism as the state $\dket{\rho^{\text{x}}(0)}$, 
we can compute the components $\rho^{\text{x}}_{n}(0)$ of $\dket{\rho^{\text{x}}(0)}$ from $\rho^{\text{x}}_{n}(0) = \trs{\rho(0)\sigma_{n}}$, where $\sigma_{n} = \{\1, \sigma_{x},\sigma_{y},\sigma_{z}\}$. Thus we get
\begin{align}
\dket{\rho^{\text{x}}(0)} = \dket{1} - \tanh[\beta \hbar \omega]\dket{x} \text{ , }
\end{align}
where we define the basis $\dket{k}=[\begin{matrix} \delta_{k1}& \delta_{kx} & \delta_{ky} & \delta_{kz} \end{matrix}]^{t}$. If we drive the system under the master equation 
\begin{align}
\rho(t) = \Lcal[\rho(t)] = \frac{1}{i\hbar} [H_{\text{x}},\rho(t)] + \gamma (t) \left[ \sigma_{z} \bullet \sigma_{z} - \bullet \right] \text{ , }
\end{align}
the superoperator $\Lmath (t)$ associated with the generator $\Lcal[\bullet]$ reads
\begin{align}
\Lmath (t) = \begin{bmatrix}
0 & 0            & 0            & 0 \\
0 & -2 \gamma(t) & 0            & 0 \\
0 & 0            & -2 \gamma(t) & -2 \omega \\
0 & 0            & 2 \omega & 0 
\end{bmatrix} \text{ . }
\end{align}
Thus, it is possible to show that the set $\{\dket{1},\dket{x}\}$ satisfies the eigenvalue equation for $\Lmath (t)$ as
\begin{align}
\Lmath (t) \dket{1} = 0 \text{ \ \ , \ \ } \Lmath (t)\dket{x} = -2\gamma(t)\dket{x} \text{ . } \label{EigenAp}
\end{align}
It can be shown that this eigenstates are nondegenerate.
Therefore, if the dynamics is adiabatic, we can write the evolved state as $\dket{\rho^{\text{x}}(t)} = c_{1}(t)\dket{1} + c_{x}(t)\dket{x}$, where $c_{y}(t) =c_{y}(0) = 0$ and $c_{z}(t)=c_{z}(0)=0$ because the coefficients evolve independently form each other. Thus, from the adiabatic solution in open quantum system given in Eq.~\eqref{AdEvolvedAp},
we obtain $c_{1}(t) = 1$ and $c_{x}(0)=-\tanh[\beta \hbar \omega]$, so that we can use $\tilde{\lambda}_{1}=0$ and $\tilde{\lambda}_{x}=-2\gamma(t)$ to obtain
\begin{align}
\dket{\rho^{\text{x}}(t)} = \dket{1} - e^{-2 \int_{0}^{t}\gamma(\xi)d\xi}\tanh[\beta \hbar \omega] \dket{x} \text{ . } \label{Aprhox}
\end{align}
Notice that Eq.~(7) in the main text directly follows by rewriting Eq.~(\ref{Aprhox}) in the standard operator formalism.
Moreover, by using this formalism, it is also possible to show that the dephasing channel can be used as a thermalization process if we suitably choose the parameter 
$\gamma(t)$ and the total evolution time $\tau_{\text{dec}}$. In fact, we can define a new inverse temperature $\beta_{\text{deph}}$ so that 
Eq.~(\ref{Aprhox}) behaves as thermal state, namely,
\begin{align}
\dket{\rho^{\text{x}}(t)} = \dket{1} - \tanh[\beta_{\text{deph}} \hbar \omega] \dket{x} \text{ . } \label{ThermalAp}
\end{align}
where we immediately identify
\begin{align}
\beta_{\text{deph}} = \frac{1}{\hbar \omega} \text{arctanh}\left[ e^{-2 \int_{0}^{t}\gamma(\xi)d\xi}\tanh(\beta \hbar \omega) \right] \text{ . }
\end{align}
In particular, by using the mean value theorem, there is a value $\bar{\gamma}$ so that $\bar{\gamma} = (1/\tau_{\text{dec}})\int_{0}^{\tau_{\text{dec}}} \gamma(t)dt$. Then, the above equation becomes
\begin{align}
\beta_{\text{deph}} = \frac{1}{\hbar \omega} \text{arctanh}\left[ e^{-2 \bar{\gamma} \tau_{\text{dec}}}\tanh(\beta \hbar \omega) \right] \text{ . }
\end{align}

In addition, heat can be computed from Eq.~\eqref{QApAd} as
\begin{align}
dQ^{\text{ad}} &= \frac{1}{D}\sum \nolimits _{i,k_{i}} c^{(k_{i})}_{i} e^{\int_{0}^{t} \tilde{\lambda}_{i,k_{i}}(t^{\prime})dt^{\prime}}\dinterpro{h(t)}{\Lmath (t)|\Dcal^{(k_{i})}_{i}(t)}dt \nonumber \\ &= \frac{1}{2} \left[ c_{1}\dinterpro{h(t)}{\Lmath (t)|1} + c_{x}e^{-2\int_{0}^{t} \gamma(t^{\prime})dt^{\prime}}\dinterpro{h(t)}{\Lmath (t)|x}\right]dt \text{ , }
\end{align}
where we already used $c_{i} = 0$, for $i = y,z$. Now, we can use that the vector $\dbra{h(t)}$ has components $h_{n}(t)$ given by $h_{n}(t) = \trs{\rho(0)H(t)}$, in which $H(t)$ is the Hamiltonian that acts on the system during the non-unitary dynamics. In conclusion, by using this result and Eq.~(\ref{EigenAp}), we get
\begin{align}
d Q^{\mathrm{ad}}(t) = 2 \hbar \omega  \tanh[\beta \hbar \omega]\gamma(t) e^{-2 \int_{0}^{t}\gamma(\xi)d\xi} dt \text{ . }
\end{align}

Now, let us to use the mean-value theorem for real functions to write $\bar{\gamma} = (1/\Delta t) \int_{0}^{t}\gamma(\xi)d\xi$ within the interval $\Delta t$, so that we get $e^{-2 \int_{0}^{t}\gamma(\xi)d\xi}=e^{-2 \bar{\gamma} \Delta t }$. It shows that the higher the mean-value of $\gamma(t)$ the faster the heat exchange ends. Now, by integrating the above result
\begin{align}
\Delta Q(\tau_{\text{dec}}) &= \int_{0}^{\tau_{\text{dec}}}d Q^{\mathrm{ad}}(t) \nonumber \\ &= 2 \hbar\omega  \tanh[\beta \hbar \omega] \int_{0}^{\tau_{\text{dec}}} \gamma(t) e^{-2 \int_{0}^{t}\gamma(\xi)d\xi} dt \text{ . } \label{InQAp}
\end{align}
To solve the above equation, we need to solve
\begin{align}
F(t) = \int_{0}^{\tau}\gamma(t)e^{-2 \int_{0}^{t}\gamma(\xi)d\xi} dt\text{ , } \label{AppF}
\end{align}
where we can note that
\begin{align}
\frac{d}{dt}\left[e^{-2\int_{t_{0}}^{t} \gamma(\xi)d\xi}\right] &= e^{-2\int_{t_{0}}^{t} \gamma(\xi)d\xi}\frac{d}{dt}\left[-2\int_{t_{0}}^{t} \gamma(\xi)d\xi\right] 
\nonumber \\ &= -2\gamma(t)e^{-2\int_{t_{0}}^{t} \gamma(\xi)d\xi} \text{ . }
\end{align}
Therefore, we can write the Eq.~\eqref{AppF} as
\begin{align}
F(t) &= -\frac{1}{2}\int_{0}^{\tau}  \frac{d}{dt}\left[e^{-2\int_{t_{0}}^{t} \gamma(\xi)d\xi}\right] dt = -\frac{1}{2} \left[e^{-2\int_{t_{0}}^{\tau} \gamma(t)dt} - 1\right] \nonumber \\ &= -\frac{1}{2} \left[e^{-2(\tau-t_{0}) \bar{\gamma}} - 1\right] \text{ . }
\end{align}
where we used the mean-value theorem in the last step. Therefore, by using this result in Eq.~\eqref{InQAp}, we find
\begin{align}
\Delta Q(\tau_{\text{dec}}) = \hbar \omega \tanh[\beta \hbar \omega] \left( 1 - e^{-2 \bar{\gamma} \tau_{\text{dec}}}\right) \text{ . }
\end{align}

In order to study the the average power for extracting/introducing the amount $|\Delta Q(\tau_{\text{dec}})|$, we define the quantity $\bar{\Pcal}(\tau_{\text{dec}}) = |\Delta Q(\tau_{\text{dec}})|/\tau_{\text{dec}}$, where $\tau_{\text{dec}}$ is the time interval necessary to extract/introduce the amount of heat $|Q(\tau_{\text{dec}})|$. Thus, from the above equation we obtain 
\begin{align}
\bar{\Pcal}(\tau_{\text{dec}}) = |\Delta Q_{\text{max}}|\eta(\tau_{\text{dec}},\bar{\gamma}) \text{ , }
\end{align}	
with $\Delta Q_{\text{max}} = \hbar\omega \tanh[\beta \hbar \omega] $ and $\eta(\tau_{\text{dec}},\bar{\gamma}) = ( 1 - e^{-2 \bar{\gamma} \tau_{\text{dec}}} )/\tau_{\text{dec}}$. This result is illustrated in Fig.~\ref{Graph2}, where we have plotted $\bar{\Pcal}(\tau_{\text{dec}})$ during the entire heat exchange (within the interval $\tau_{\text{dec}}$) as a function of $\tau_{\text{dec}}$.  Notice that, as in the case of $\Delta Q(\tau_{\text{dec}})$, the asymptotic behavior of the average power is independent of $\gamma_0$.

For our dynamics, the entropy variation is obtained from Eq.~\eqref{EntropyProAp} for a one-dimensional block Jordan decomposition. Thus, by computing $\dbra{\rho^{\text{Ad}}_{\log}(t)}$, where we find 
\begin{align}
\dbra{\rho^{\text{Ad}}_{\log}(t)} &= \log \left(\frac{1 - g^2(t)}{4} \right) \dbra{1} -2 \text{arctanh} [g(t)] \dbra{x} \text{ , }
\end{align}
with $g(t)\!=\!e^{-2 \int_{0}^{t}\gamma(\xi)d\xi}\tanh(\beta \hbar \omega)$. Then, from Eq.~\eqref{EntropyProAp} we get
\begin{align}
\dot{S}(t) &=\frac{1}{2}\sum \nolimits _{i = {0}}^{1} c_{i} e^{\int_{0}^{t} \tilde{\lambda}_{i}(t^{\prime})dt^{\prime}}
\Gamma_{i}(t) \text{ , }
\end{align}
where $\Gamma_{i}(t) = \lambda_{i}(t) \dinterpro{\rho^{\text{ad}}_{\log}(t)}{\Dcal_{i}(t)}$. Hence, from the set of adopted values for our parameters and the spectrum of the Lindbladian, we get
\begin{align}
\dot{S}(t) = 4 g(t) \gamma(t) \text{arctanh} [g(t)] \text{ . }
\end{align}

\vspace{0.5cm}

\noindent\textbf{Trapped-ion experimental setup}

\vspace{0.3cm}


We encode a qubit into hyperfine energy levels of a trapped Ytterbium ion $^{171}$Yb$^+$, denoting its associated states by $\ket{0} \equiv \ket{^{2}S_{1/2}; F=0, m=0}$ and $\ket{1} \equiv \ket{^{2}S_{1/2}; F=1, m=0}$. By using an arbitrary waveform generator (AWG) we can drive the qubit through either a unitary or a non-unitary dynamics (via a frequency mixing scheme). The detection of the ion state is obtained from use of a ``readout" laser with wavelength $369.526$ nm.

Applying a static magnetic field with intensity $6.40$~G, we get a frequency transition between the qubit states given by $\omega_{\text{hf}} = 2\pi \times 12.642825$~GHz. Therefore, by denoting the states $\ket{0}$ and $\ket{1}$ as ground and excited states, respectively, the inner system Hamiltonian is given by
\begin{align}
H_{0} = \frac{\hbar \omega_{\text{hf}}}{2} \sigma_{z}
\end{align}
where $\sigma_{z} = \ket{1}\bra{1} - \ket{0}\bra{0}$. Therefore, to unitarily drive the system through coherent population inversions within the subspace $\{\ket{0},\ket{1}\}$, we use a microwave at frequency $\omega_{\text{mw}}$ whose magnetic field
\begin{align}
\vec{B}_{\text{un}}(t) = \vec{B}_0 \cos \omega_{\text{mw}} t
\end{align}
interacts with the electron magnetic dipole moment $\hat{\mu} = \mu_{M}\hat{S}$, with $\mu_{M}$ a constant and $\hat{S}$ is the electronic spin. Then, the
system Hamiltonian reads
\begin{align}
H(t) = H_{0} - \hat{\mu} \cdot \vec{B}_{\text{un}}(t) .
\end{align}
Thus, by defining the Rabi frequency $\hbar\Omega_{\text{R}} \equiv - \mu_{M}|\vec{B}_0|/4$~\cite{Wineland:98}, we obtain that the effective Hamiltonian that drives the qubit is (in interaction picture)
\begin{align}
H_{I}(t) = \frac{\hbar \omega}{2} \sigma_{z} + \frac{\hbar \Omega_{\text{R}}}{2} \sigma_{x} \text{ , }
\end{align}
where $\omega = \omega_{\text{hf}} - \omega_{\text{mw}}$ and $\sigma_{x} = \ket{1}\bra{0} + \ket{0}\bra{1}$. By using the AWG we can efficiently control the parameters $\omega$ and $\Omega_{\text{R}}$. In particular, in our experiment to implement the Hamiltonian $\tilde{H}_{\text{x}}$, we have used a resonant ($\omega_{\text{mw}}=\omega_{\text{hf}}$) microwave with Rabi frequency $\Omega_{\text{R}} = \tilde{\omega}$, while the frequency $\omega_{\text{hf}}$ has been adjusted around $2\pi \times 12.642$~GHz, with $\tilde{\omega}$ modulated by using the channel 1 (CH1) of the AWG.

After the experimental qubit operation, we use the state-dependent florescence detection method to implement the quantum state binary measurement. We can observe on average 13 photons for the bright state $\ket{1}$ and zero photon for the dark state $\ket{0}$ in the 500 $\mu s$ detection time interval, as shown in Fig.~\ref{Fig_histgram}. These scattered photons at $396.526$ nm are collected by an objective lens with numerical aperture $NA=0.4$. After the capture of these photons, they go through an optical bandpass filter and a pinhole, after which they are finally detected by a photomultiplier tube (PMT) with $20\%$ quantum efficiency. By using this procedure, the measurement fidelity is measured to be $99.4\%$.

Due to the long coherence time of the hyperfine qubit, the decoherence effects can be neglected in our experimental timescale. However, since we are interested in a nontrivial non-unitary evolution, we need to perform environment engineering. This task can be achieved by using a Gaussian noise source to mix the carrier microwave
$\vec{B}_{\text{un}}(t)$ by a frequency modulation (FM) method.
Thus, by considering the noise source encoded in the function $\eta(t)=A g(t)$, where $A$ is average amplitude of the noise and $g(t)$ is a random
analog voltage signal, the driving magnetic field will be in form
\begin{align}
\vec{B}_{\text{n-un}}(t) = \vec{B}_0 \cos [\omega t + C\eta(t) \, t ]
\end{align}
where $|\vec{B}_0|$ is field intensity and $C$ is the modulation depth supported by the commercial microwave generator E8257D. If $C$ is a fixed parameter (for example, $C = 96.00$~KHz/V), the dephasing rate $\gamma(t)$ associated with Lindblad equation
\begin{align}
\dot{\rho}(t) = \frac{1}{i\hbar} [\tilde{H}_{\text{x}},\rho(t)] + \gamma(t) \left[\sigma_z \rho(t) \sigma_z - \rho(t)\right] \text{ , }
\label{Lindbald}
\end{align}
is controlled from the average amplitude of the Gaussian noise function $\eta(t)$. To see that $\eta(t)$ is a Gaussian function in the frequency domain,
we show its spectrum in Fig. \ref{Fig_spetrum}.

In order to certify that the decoherence channel is indeed a $\sigma_z$ channel (dephasing channel) in our experiment, we employed
quantum process tomography. A general quantum evolution can be typically
described by the operator-sum representation associated to a trace-preserving map $\varepsilon$. For an arbitrary input state $\rho$, the output state
$\varepsilon(\rho)$ can be written as~\cite{Nielsen:Book}
\begin{align}
\varepsilon (\rho ) = \sum\limits_{m,n} {{\chi _{mn}}A_m\rho {A^\dag_n }} \text{ , }
\end{align}
where $A_m$ are basis elements (usually a fixed reference basis) that span the state space associated with $\rho$ and $\chi _{mn}$ is the matrix element of the so-called {\it process matrix} $\chi$, which can be measured by  quantum state tomography. In a single qubit system, we take $A_0$ = $I$, $A_1$ = $\sigma_x$, $A_2$ = $\sigma_y$, $A_3$ = $\sigma_z$. The quantum process tomography is carried out for the quantum process described by the Lindblad equation given by Eq.~\eqref{Lindbald},
where $H(t)=\omega \sigma_x$, with $ \omega = 5.0\times2\pi$ KHz and $\gamma=2.5$ KHz. We fixed the total evolution time as 0.24 ms (here, the noise amplitude is 1.62 V and the modulation depth is 96.00 KHz). The resulting estimated process matrix is shown in Fig. \ref{Fig_process}. We can calculate the fidelity between the experimental process matrix $\chi_{\text{exp}}$ and the theoretical process matrix $\chi_{\text{id}}$
\begin{align}
\Fcal({\chi _{\text{exp} }},{\chi _{\text{id}}}) = \left[\text{Tr}\sqrt{ \sqrt{\chi _{\text{exp} }} \text{ } \chi _{\text{id}} \sqrt{\chi _{\text{exp} }} }\right]^2
\end{align}
We measured several process with different evolution times. For example, when the amplitude of the noise is set to 1.54V, the process fidelities are measured as  $\Fcal_{t_{1}} = 99.27 \%$, $\Fcal_{t_{2}} = 99.50 \%$, $\Fcal_{t_{3}} =  99.72 \%$, $\Fcal_{t_{4}} = 99.86 \%$ and $\Fcal_{t_{5}} = 99.87 \%$, at times $t_{1} = 0.08$ ms, $t_{2} =0.16$ ms, $t_{3} = 0.24$ ms, $t_{4} = 0.32$ ms and $t_{5} = 0.40$ ms, respectively. Thus, the dephasing channel can be precisely controlled as desired and it can support the scheme to implement the time-dependent dephsing in experiment.

The function $\eta(t)$ depends on an amplitude parameter $A$, which is used to control $\gamma (t)$. As shown in Fig. \ref{Fig_dephsaing}, we experimentally measured the relation between $A$ and $\gamma (t)$ for a situation where $\gamma(t)$ is a time-independent value $\gamma_{0}$. As result, we find a linear relation
between $\sqrt{\gamma_{0}}$ and $A$, which reads
\begin{align}
\sqrt{\gamma_{0}} = 29.81 A + 1.74 \text{ . }
\end{align}
For the case $A = 0$, we get the natural dephasing rate $\gamma_{\text{nd}} = 1.74^2$~Hz of the physical system. Thus, we can see that, if we change the parameter $A$, which we can do with high controllability, the quantity $\sqrt{\gamma_{0}}$ can be efficiently controlled. On the other hand, if we need a time-dependent rate $\gamma(t)$, we just need to consider a way to vary $A$ as a function $A(t)$. To this end, we use a second channel (CH2) of the AWG to perform amplitude modulation (AM) of the Gaussian noise. The temporal dependence of $A(t)$ is achieved by programming the channel (CH2) to change during the evolution time.


In order to guarantee that the dynamics of the system is really adiabatic \cite{Sarandy:05-1} we compute the fidelity $\Fcal(\tau_{\text{dec}})$
of finding the system in a path given by Eq.~({\color{blue}5}), where $\Fcal(t)=\text{Tr}\{[\rho_{\text{exp}}^{1/2}(t)\rho_{\text{ad}}(t)\rho_{\text{exp}}^{1/2}(t)]^{1/2}\}$, with $\rho_{\text{ad}}(t)$ the density matrix provided Eq.~({\color{blue}5}) and $\rho_{\text{exp}}(t)$ the experimental density matrix obtained from quantum tomography. In  Table~\ref{Table} we show the minimum experimental fidelity $\Fcal_{\text{min}} = \min_{\tau_{\text{dec}}}\Fcal(\tau_{\text{dec}})$ for several choices of the parameter $\gamma_{0}$. This result shows that the system indeed evolves as predicted by the adiabatic solution for every $\gamma_{0}$ and $\tau_{\text{dec}}$ with excellent experimental agreement.

\begin{table}[h!]
	\centering
	\caption{Minimum value of experimental fidelity $\Fcal_{\text{min}}$ for each choice of $\gamma_{0}$. The maximum experimental error $\Delta\Fcal_{\text{min}}$ for $\Fcal_{\text{min}}$ is about $\Delta\Fcal_{\text{min}} = 0.13\%$ of $\Fcal_{\text{min}}$.}
	\label{Table}
  \begin{center}
\begin{tabular}{cccccc}
	\multicolumn{1}{l|}{$\gamma_{0}$}         & 314 Hz    & 628 Hz    & 1257 Hz   & 3142 Hz   & 6283 Hz \\ \hline
	\multicolumn{1}{l|}{$\Fcal_{\text{min}}$} & 0.9971(3) & 0.9965(4) & 0.9980(7) & 0.9952(8) & 0.9942(9)
\end{tabular}
  \end{center}
\end{table}

\section*{DATA AVAILABILITY} The data that support the findings of this study are available from the corresponding authors upon reasonable request.

\section*{CODE AVAILABILITY} The code that support the findings of this study are available from the corresponding authors upon reasonable request.

\section*{Acknowledgment} We thank Yuan-Yuan Zhao, Zhibo Hou, Jun-Feng Tang, and Yue-xin Huang for valuable discussion. This work was supported by the National Key Research and Development Program of China (No. 2017YFA0304100), National Natural Science Foundation of China (Nos. 61327901, 61490711, 11774335, 11734015), Anhui Initiative in Quantum Information Technologies (AHY070000, AHY020100), Anhui Provincial Natural Science Foundation (No. 1608085QA22), Key Research Program of Frontier Sciences, CAS (No. QYZDY-SSWSLH003), the Fundamental Research Funds for the Central Universities (WK2470000026), and the China Postdoctoral Science Foundation (Grant No. 2020M671861). A.C.S. is supported by Conselho Nacional de Desenvolvimento Cient\'{\i}fico e Tecnol\'ogico (CNPq-Brazil). M.S.S. is supported by CNPq-Brazil (No. 303070/2016-1) and Funda\c{c}\~ao Carlos Chagas Filho de Amparo \`a Pesquisa do Estado do Rio de Janeiro (FAPERJ) (No. 203036/2016). D.O.S.P is supported by Brazilian funding agencies CNPq (Grants No. 142350/2017-6 and 305201/2016-6), and FAPESP (Grant No. 2017/03727-0). A.C.S., D. O. S.-P. and M.S.S. also acknowledge financial support in part by the Coordena\c{c}\~ao de Aperfei\c{c}oamento de Pessoal de N\'{\i}vel Superior - Brasil (CAPES) (Finance Code 001)
and by the Brazilian National Institute for Science and Technology of Quantum Information [CNPq INCT-IQ (465469/2014-0)]. A.C.S., D. O. S.-P. and M.S.S. would like to thank F. Brito and J. G. Filgueiras for fruitful comments.

\section*{Author Contributions} A.C.S., D.O.S.-P. and M.S.S. developed and performed the theoretical analysis.
C.-K.H., J.-M.C., Y.-F.H. and C.-F.L. designed the experiment.
C.-K.H., J.-M.C. and Y.-F.H. performed the experiment.
C.-K.H., Y.-F.H., A.C.S., D.O.S.-P. and M.S.S. wrote the manuscript.
C.-F.L. and G.-C.G. supervised the project.
All authors discussed and contributed to the analysis of the experimental data.

\section*{COMPETING INTERESTS}

The authors declare no competing interests.


\newpage
\onecolumngrid
\section*{Figure Legends}

\begin{figure}[ht]
	\centering
	\includegraphics[scale=0.15]{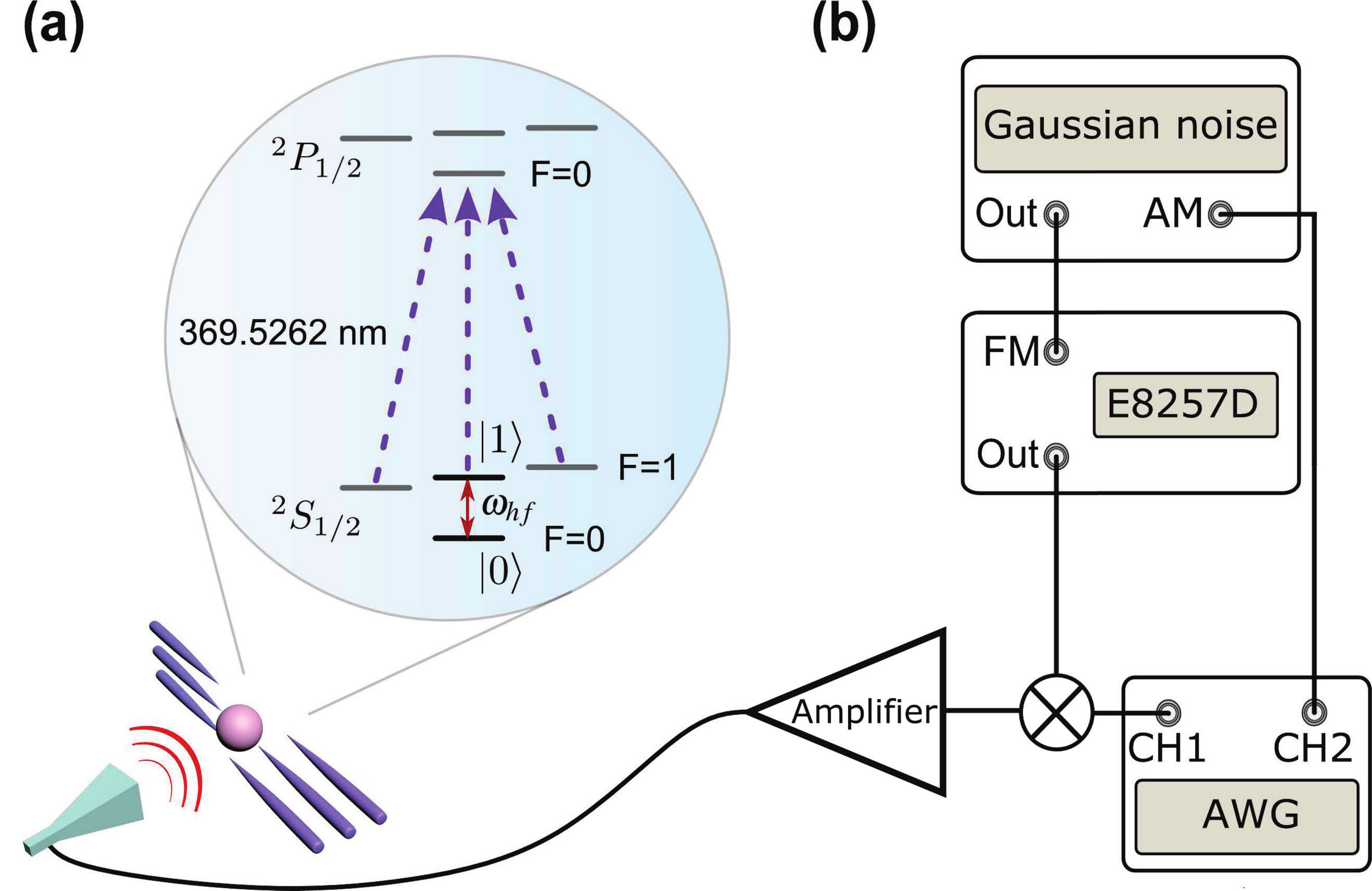}
	\caption{\textbf{Experimental scheme to investigate the thermodynamics of adiabaticity in open quantum systems.}
		(a) Schematic diagram of the six-needle Paul trap and relevant levels of the $^{171}$Yb$^+$ ion.
		(b) Experimental microwave instrument for generating the field to drive the two level system.
		The AWG is programmed to implement the target Hamiltonian and control the amplitude of the Gaussian noise which is used as a dephasing channel.}
	\label{Fig1}
\end{figure}

\begin{figure}[ht]
	\centering
	\includegraphics[scale=0.33]{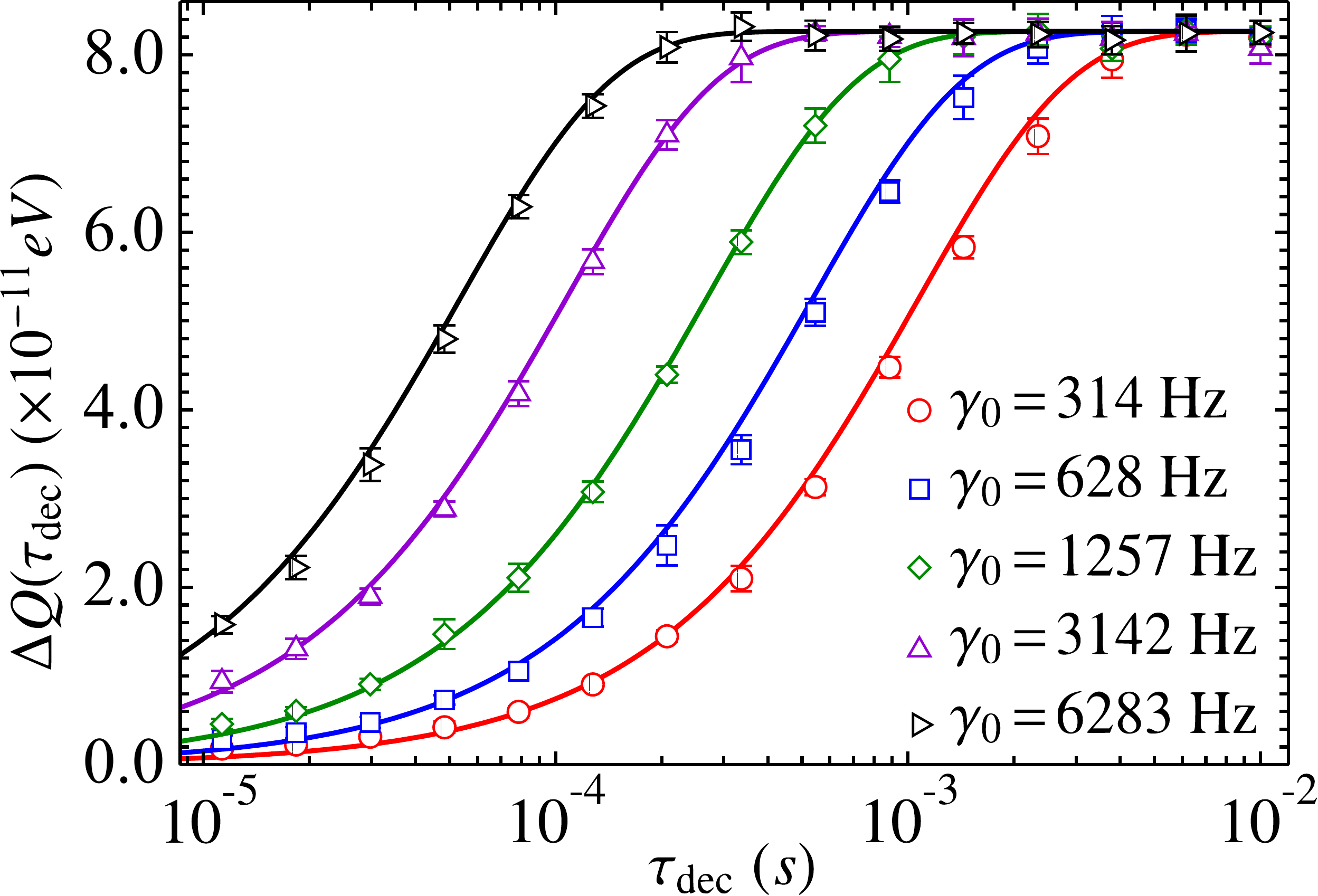}
	\caption{\textbf{Heat $\Delta Q(\tau_{\text{dec}})$ as a function of the total evolution time $\tau_\text{dec}$ for several values of the parameter $\gamma_{0}$.} We use $\hbar\omega = 82.662$~peV and $\beta^{-1} = 17.238$~peV, with the physical constants $\hbar \approx 6.578 \cdot 10^{-16}$~eV$\cdot$s and $k_{\text{B}} \approx 8.619 \cdot 10^{-5}$~eV/K~\cite{Mohr:16}.}
	\label{Graph1}
\end{figure}

\begin{figure}[t]
	\centering
	\includegraphics[scale=0.33]{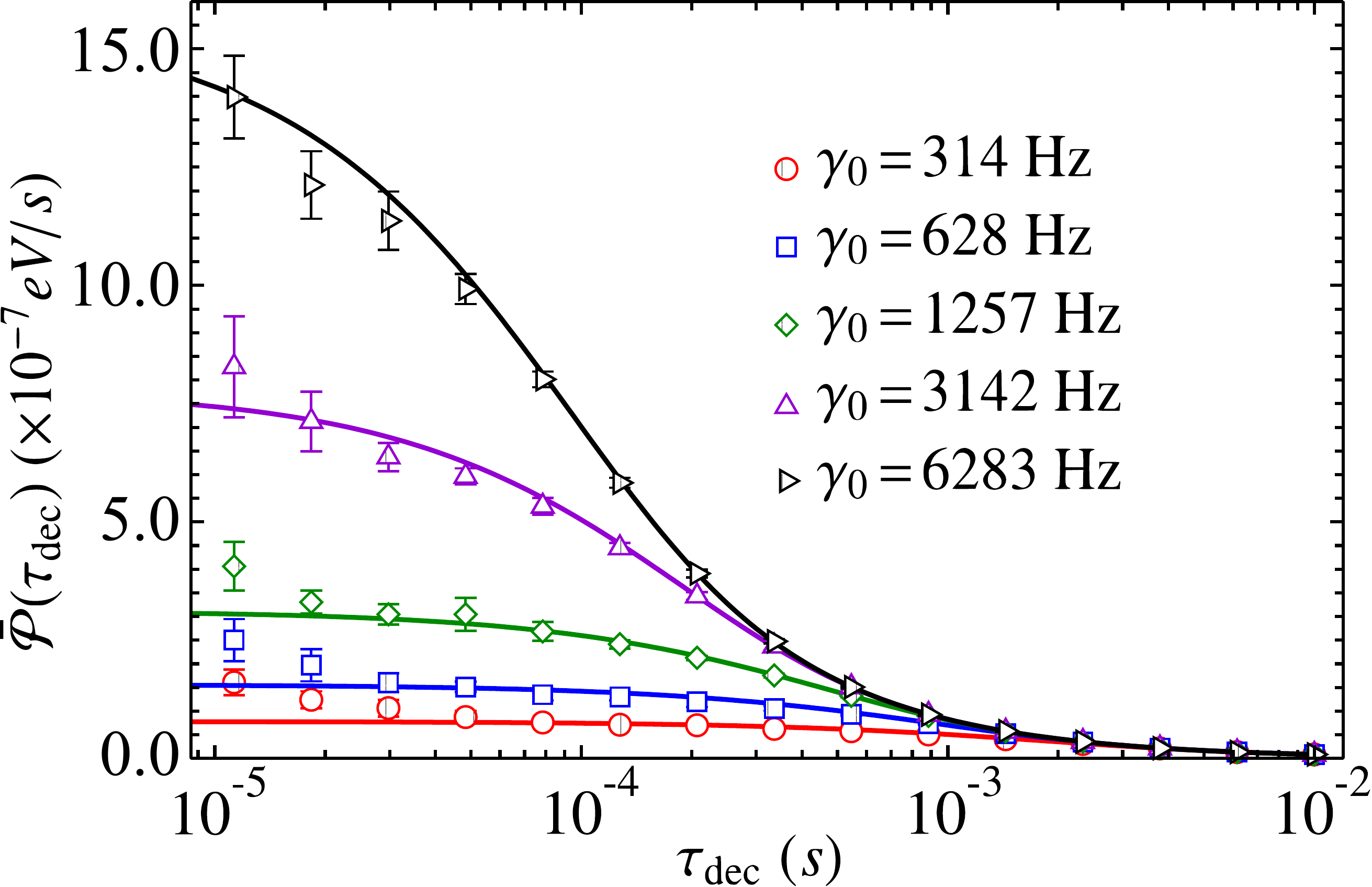}
	\caption{\textbf{Average power $\bar{\Pcal}(\tau_{\text{dec}})$ as a function of $\tau_\text{dec}$ for several values of $\gamma_{0}$.} Here we use $\hbar\omega = 82.662$~peV and $\beta^{-1} = 17.238$~peV, with the physical constants $\hbar \approx 6.578 \cdot 10^{-16}$~eV and $k_{\text{B}} \approx 8.619 \cdot 10^{-5}$~eV~\cite{Mohr:16}.}
	\label{Graph2}
\end{figure}

\begin{figure}[t]
	\centering
	\includegraphics[scale = 0.55]{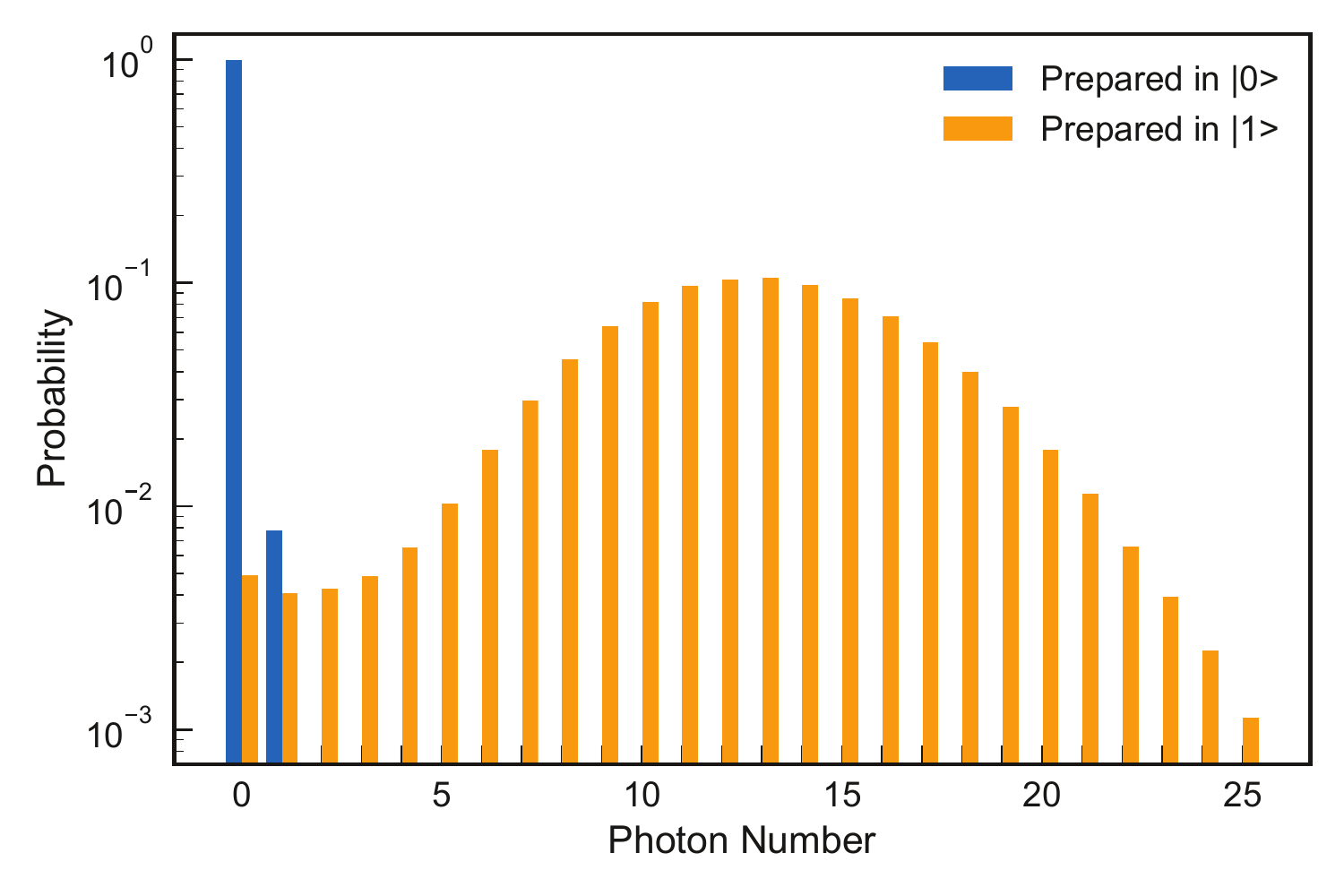}
	\caption{\textbf{Histograms of detected photons after the ion is prepared in $\ket{0}$ and $\ket{1}$.} All data is obtained under 100 000 measurement repetitions.}
	\label{Fig_histgram}
\end{figure}

\begin{figure}[t]
	\centering
	\includegraphics[scale = 0.55]{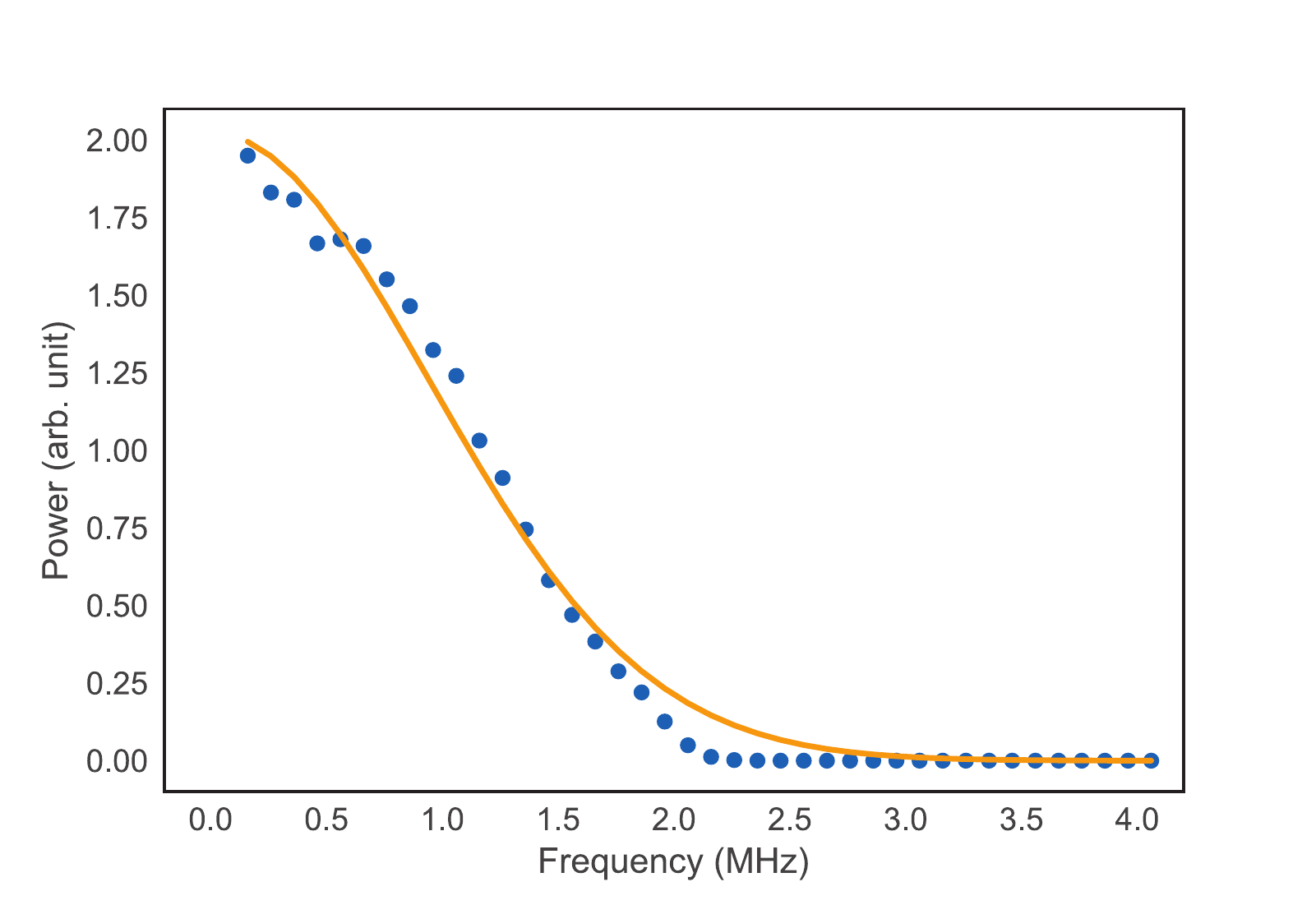}
	\caption{\textbf{Spectrum of the noise source.} The noise source is provided by the commercial microwave generator E8257D. Dots are measured data and the solid curve is a Gaussian fit to the data.}
	\label{Fig_spetrum}
\end{figure}

\begin{figure}[t]
	\centering
	\includegraphics[scale = 0.4]{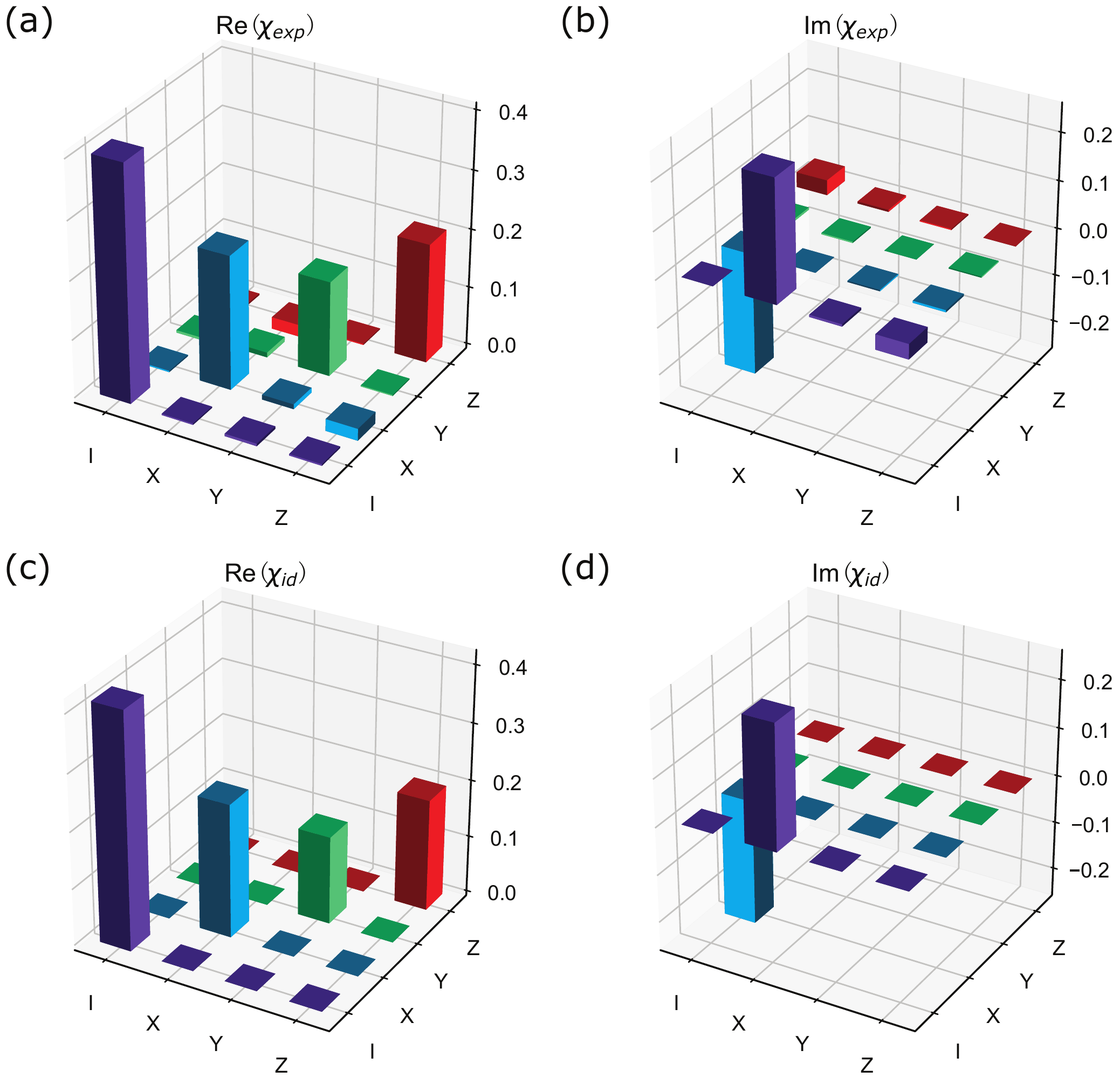}
	\caption{\textbf{Process matrix via process tomography.} The plots (a) and (b) are the real and imaginary parts of $\chi$
		obtained from the experimental measured data. Plots (c) and (d) are the real and imaginary parts of $\chi$ given by numerical simulation.	}
	\label{Fig_process}
\end{figure}

\begin{figure}[t]
	\centering
	\includegraphics[scale = 0.5]{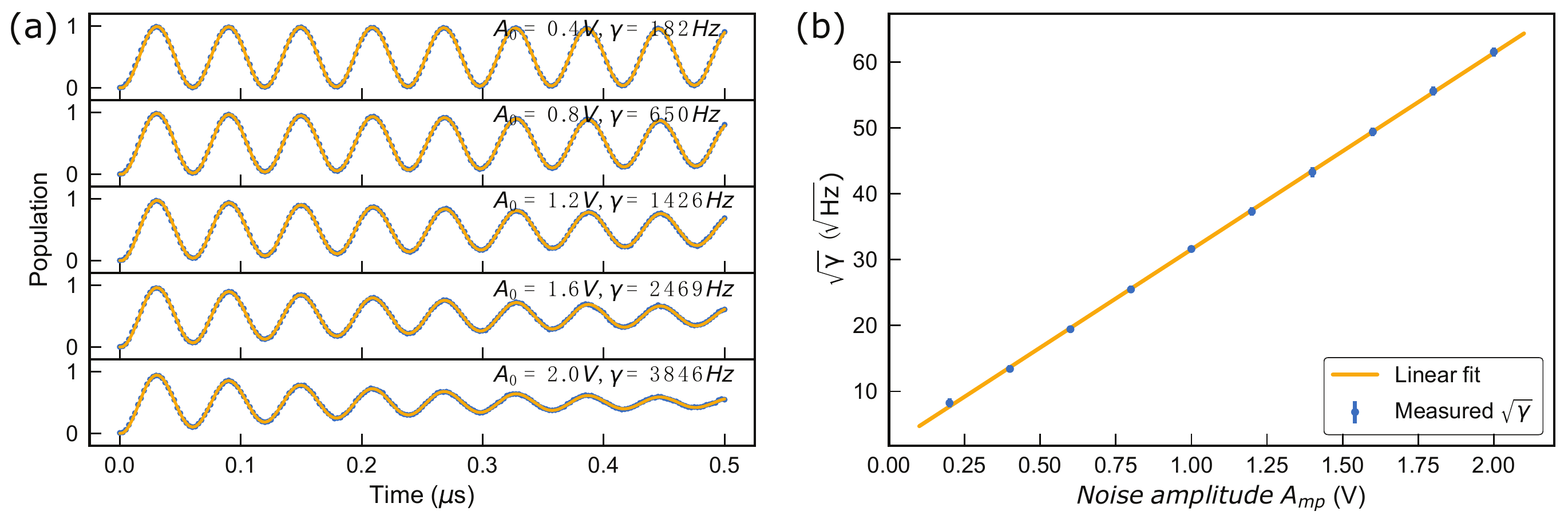}
	\caption{\textbf{Dephasing rate controlled by the amplitude of noise, here $C$ is fixed as $C = 96.00$~KHz/V.} {\color{blue}(a)} Rabi oscillations between states $\ket{0}$ and $\ket{1}$ under  different noise intensities. From top to bottom, the noise amplitude is set to 0.4 V, 0.8 V, 1.2 V, 1.6 V and 2.0 V, with the corresponding damping rates 182 Hz, 650 Hz, 1426 Hz, 2469 Hz and 3846 Hz, respectively. {\color{blue}(b)} Dephasing rate as a function of the noise amplitude. Points are measured data. A linear fit is obtained. Without driving noise (noise amplitude is zero), the dephasing rate of the qubit is fitted as 3.03 Hz, which is caused by the magnetic fluctuation in the laboratory.
	}
	\label{Fig_dephsaing}
\end{figure}

\newpage

\end{document}